\documentclass[11pt]{iopart}
\usepackage{iopams}
\usepackage{graphicx}
\begin{document}
\title[Finite-Size Scaling Study of Aging during Coarsening in Non-Conserved Ising Model]
{Finite-Size Scaling Study of Aging during Coarsening in 
Non-Conserved Ising Model: The case of zero temperature quench}
\author{ Nalina Vadakkayil, Saikat Chakraborty and Subir K. Das}
\address{Theoretical Sciences Unit, Jawaharlal Nehru Centre for Advanced Scientific Research,
Jakkur P.O., Bangalore 560064, India.}
\ead{das@jncasr.ac.in}
\vspace{10pt}
\begin{indented}
\item[]June 2018
\end{indented}
\begin{abstract}
\par
Following quenches from random initial configurations to zero temperature, we study aging during evolution 
of the ferromagnetic (nonconserved) Ising model towards equilibrium, via Monte Carlo simulations of very large 
systems, in space dimensions $d=2$ and $3$. 
Results for the two-time autocorrelations, obtained by using different acceptance probabilities for the spin-flip trial 
moves, are in agreement with each other. 
We demonstrate the scaling of this quantity with respect to $\ell/\ell_w$, where $\ell$ and $\ell_w$ 
are the average domain sizes at $t$ and $t_w$ $(\leqslant t)$,  
the observation and waiting times, respectively. The scaling functions are shown to be of power-law type for 
$\ell/\ell_{w} \rightarrow \infty$. The exponents of these power-laws have been estimated via the finite-size 
scaling analyses and discussed with reference to the available results from non-zero  temperatures. While in $d=2$ we 
do not observe any temperature dependence, in the case of $d=3$ the outcome for quench to zero temperature is 
very different from the available results for high temperature and violates a lower bound, which we explain 
via structural consideration. We also present results on the freezing phenomena that this 
model exhibits at zero temperature. Furthermore, from simulations of extremely large system, thereby 
avoiding the freezing effect, it has been confirmed that 
the growth of average domain size in $d=3$, that remained a puzzle in the literature, follows the Lifshitz-Allen-Cahn law 
in the asymptotic limit.
\end{abstract}
\vspace{2pc}
\maketitle
\section{Introduction}
Following quench from a homogeneous configuration to a state inside the coexistence curve, as a system evolves 
towards the new equilibrium, various structural quantities exhibit interesting scaling 
properties \cite{bray,puri,dfish,liu,satya,yeu,cor,hen,ohta,aren,lor}. 
In this context, a rather general order-parameter correlation function is defined by connecting two space 
points ($\vec{r}_{1},\vec{r}_2$) and two times $(t,t_{w})$, and is written as \cite{puri}
\begin{eqnarray}\label{e5_corfn}
C_{22}(\vec{r}_{1},\vec{r}_2;t,t_{w})=&\langle {\psi(\vec{r}_{1},t)\psi(\vec{r}_{2},t_{w})}\rangle-\nonumber \\
&\langle{\psi(\vec{r}_{1},t)}\rangle \langle{\psi(\vec{r}_{2},t_{w})}\rangle.
\end{eqnarray}
Here $\psi$ is a space- and time-dependent order-parameter field. 
For isotropic structures, which we assume to be true for the cases addressed in this paper, the space dependence 
in $C_{22}$ comes through $r=|\vec{r}_{1}-\vec{r}_{2}|$, the scalar distance between $\vec{r}_1$ and 
$\vec{r}_2$. For $t=t_{w}$, $C_{22}$, to be denoted by $C(r,t)$, is referred to as the two-point equal-time correlation 
function \cite{bray,puri}. On the other hand, for $\vec{r}_{1}=\vec{r}_{2}$, we call $C_{22}$ 
the two-time autocorrelation function \cite{puri}. The latter quantity, 
that will be represented by $C_{\textrm{ag}}(t,t_{w})$, is often used for studying aging in nonequilibrium 
systems \cite{puri,dfish}, where $t_{w}$ $(\leq t)$ is referred to as the waiting time or the age of the system. 
It is worth mentioning here that $C_{\textrm{ag}}(t,t_w)$ may contain information on relaxation related to equilibration  
inside individual domains as well.
\par
The two-point equal-time correlation function typically exhibits the scaling behavior \cite{bray,puri,ohta}
\begin{eqnarray}\label{e5_sclcor}
 C(r,t) \equiv \tilde{C} (r/\ell(t)),
\end{eqnarray}
where $\tilde{C}$ is a time independent master function \cite{bray} and $\ell$ is the average length of domains that are  
rich in particles or spins of one or the other type. Usually, $\ell$ grows in a power-law manner \cite{bray}, 
with exponent $\alpha$, as 
\begin{equation}\label{e_growth}
\ell \sim t^{\alpha}. 
\end{equation}
The scaling property in Eq. (\ref{e5_sclcor}), valid for non-fractal morphology, implies self-similarity, viz., 
the structures at two different times differ 
from each other only by a change in the length scale \cite{bray}. 
On the other hand, $C_{\textrm{ag}}(t,t_{w})$, in many situations, exhibits the scaling 
form \cite{puri,dfish,liu,yeu,cor,hen,lor,mid1,mid2}
\begin{equation}\label{e5_sclage}
 C_{\textrm{ag}}(t,t_{w}) \equiv \tilde{C}_{\textrm{ag}}(x);~x=\frac{\ell}{\ell_{w}},
\end{equation}
where $\ell_w$ is the characteristic length scale of the system at time $t_w$. 
\par
There has been serious interest in understanding the forms of these correlation functions for coarsening 
dynamics with and without conservation \cite{bray,puri} of the total value of the order parameter 
($={\int_{V}} d\vec{r} \psi(\vec{r},t)$, $V$ being the system volume). Remarkable progress has been made 
with respect to the nonconserved dynamics \cite{bray,puri}, for scalar as well as vector order 
parameters. A large fraction of the studies in the nonconserved variety are related to the coarsening 
in ferromagnetic Ising model \cite{bray,puri} ($\langle ij \rangle$ stands for nearest neighbors)
\begin{equation}\label{e5_ising}
 H=-J\sum_{\langle ij \rangle} {S_{i}S_{j}},~S_{i}=\pm 1,~ J>0, 
\end{equation}
or in the time-dependent Ginzburg-Landau (TDGL) model \cite{bray,puri}, the latter being essentially a 
coarse-grained version of the kinetic Ising model. 
\par
Ohta, Jasnow and Kawasaki (OJK) \cite{ohta}, via a Gaussian approximation of an auxiliary field \cite{bray,puri,ohta}, 
obtained an expression for $C_{22}$ in the case of nonconserved scalar order-parameter. This reads
\begin{equation}\label{e5_ojk_general}
C_{22}(r;t,t_{w})=\frac{2}{\pi} \sin^{-1}{\gamma},
\end{equation}
where
\begin{equation}\label{e5_ojk_gamma}
\gamma =\bigg ({\frac{2\sqrt{tt_{w}}}{t+t_{w}}}\bigg)^{d/2} \exp {\bigg[\frac{-r^2}{4D(t+t_{w})}\bigg]},
\end{equation}
$d$ being the system dimension and $D$ a diffusion constant. For $t=t_{w}$, from Eqs. (\ref{e5_ojk_general}) 
and (\ref{e5_ojk_gamma}) one obtains
\begin{equation}\label{e5_ojk}
C(r,t)= \frac{2}{\pi} \sin^{-1} \bigg[\exp \bigg(\frac{-r^2}{8Dt} \bigg) \bigg].
\end{equation}
On the other hand, for $r=0$ and $t>>t_{w}$, we have 
\begin{equation}\label{e5_age_d}
C_{\textrm{ag}}(t,t_{w}) \sim \bigg(\frac{t}{t_w}\bigg)^{-d/4}.
\end{equation}
Given that \cite{bray,puri,all} the value of $\alpha$ is $1/2$ for the nonconserved Ising model, Eq. (\ref{e5_age_d}) 
implies
\begin{equation}\label{e5_age_lambda}
C_{\textrm{ag}}(t,t_{w}) \sim \bigg(\frac{\ell}{\ell_w}\bigg)^{-\lambda};~\lambda=\frac{d}{2}.
\end{equation}
Liu and Mazenko (LM) \cite{liu}, via somewhat similar Gaussian approximation of the auxiliary field of the order 
parameter in the TDGL equation, obtained different dimension dependence  for $\lambda$. Exact solution of the 
dynamical equation for $C_{22}$, that LM constructed, provides the result same as the OJK one in $d=1$. 
However, (approximate) solutions of the above mentioned equation in $d=2$ and $3$ provide \cite{liu}
$\lambda \simeq 1.29$ and $\simeq 1.67$, respectively.
\par
For the exponent $\lambda$, Fisher and Huse (FH) \cite{dfish} provided the bounds
\begin{equation}\label{e5_fh}
\frac{d}{2} \leq \lambda \leq d.
\end{equation} 
Notice here that the lower bound of Eq. (\ref{e5_fh}) coincides with the value quoted in Eq. (\ref{e5_age_lambda}), 
outcome of the OJK theory. Later, Yeung, Rao and Desai (YRD) \cite{yeu}, by incorporating the structural 
differences between the conserved and nonconserved dynamics, obtained a more general lower bound as 
\begin{equation}\label{e5_yrd}
\lambda \geqslant \frac{d+\beta}{2}, 
\end{equation}
where $\beta$ is a power-law exponent related to the small wave-number ($k$) enhancement of the 
structure factor \cite{yeu2,satya2}:
\begin{equation}\label{e5_small_wave}
S(k,t) \sim k^{\beta}.
\end{equation}
It has been shown that $\beta=0$ for nonconserved Ising dynamics \cite{yeu2,satya2}. This leads to the agreement 
of YRD bound with the FH lower bound. Here note that $S(k,t)$ is the Fourier transform of $C(r,t)$ and has the 
scaling form \cite{bray,puri} 
\begin{equation}\label{e5_sclsf}
S(k,t) \equiv \ell^{d} \tilde{S} (k\ell),
\end{equation}
where $\tilde {S}(k\ell)$ is a time independent master function. 
\par
Predictions of both OJK and LM follow the FH bounds. We mention here that there exists an argument, related to 
percolation, by FH \cite{dfish}, that suggests $\lambda=d-a$, where $a$ is the inverse of the exponent for the 
power-law singularity of the percolation correlation length. This, e.g., provides $\lambda=5/4$ in $d=2$. 
However, FH \cite{dfish} cautioned about using this argument, as well as their upper bound. 
\par
Monte Carlo (MC) simulations of the nonconserved Ising model in $d=2$ showed consistency \cite{mid1,das} 
with the OJK function of Eq. (\ref{e5_ojk}) and the LM value for $\lambda$. The latter fact appeared 
true \cite{mid1,das} in $d=3$ as well, for quenches to certain nonzero temperatures ($T_f$) from the initial 
temperatures ($T_i$) that are far above the critical value ($T_c$). However, the $d=3$ Ising model appears to be different 
and difficult \cite{das,amar,shore,lipow,cue,cor2,ole,ole2,chak} for $T_{f}=0$. In this case, simulation reports on the time 
dependence of $\ell$ differ from the theoretical expectation \cite{all}. While some works reported $\alpha=1/3$, 
a few reported even slower growth. In recent works \cite{cor2,chak}, it has been shown, via simulations of very 
large systems, that the (theoretically) expected value $\alpha=1/2$ becomes visible only at very late time. 
\par
Furthermore, studies with smaller systems, for $T_f=0$, revealed interesting freezing behavior with respect to 
reaching the expected ground state \cite{ole,ole2}. Unusual structural aspects were also reported for $d=3$ \cite{ole,ole2}. 
In the structural context, we showed that $\tilde{C}$, unlike the $d=2$ case, differs from that at high 
temperatures \cite{das,chak_2}. Given the connection between structural and aging properties, discussed above, 
it is then natural to ask the question: Does there exist difference in the values of $\lambda$ for $T_f=0$ and 
$T_f > 0$? Our recent letter \cite{chak_2}, in fact, suggested the violation of the FH lower bound for 
$T_f=0$ in $d=3$. To confirm that, better analysis of data are needed. Furthermore, even though all the above 
mentioned studies of kinetic Ising model use (Glauber) spin-flip \cite{lan,gla} as trial move during the MC 
simulations \cite{lan}, that does not preserve the global order parameter, these moves were accepted with different 
probabilities in different studies. For example, in our previous studies, Metropolis algorithm \cite{lan} was used, 
whereas Refs. \cite{ole} and \cite{ole2} used the Glauber algorithm \cite{lan,gla}. Thus, in addition to providing 
details related to our recent letter \cite{chak_2} and arriving 
at appropriate conclusion via more accurate analysis, we also undertake a comprehensive study 
to compare results from these two different algorithms, including results on the freezing phenomena  
that prevent the systems from reaching the ground state. Given the anomalies reported at $T_f=0$, this 
exercise, we feel, is important.
\par
In this paper, all the results are presented from $T_f=0$. Via state-of-the-art finite-size scaling 
analysis \cite{mid1,mid2,lan,fish,fis_3,heer,das2} of the MC \cite{lan} 
simulation results, we arrive at the following conclusions for the decay of $C_{\textrm{ag}}(t,t_w)$. We 
confirm that there exists no  
temperature dependence in pattern, growth and aging in the case of $d=2$. On the other hand, for $d=3$, the value 
of the aging exponent, estimated from significantly long period of simulations, indeed violates the FH 
lower bound. This, however, can be explained via the structural consideration of YRD. 
On the issue of freezing, in agreement with a previous work, we find that for $d=3$ and $T_{f}=0$ systems almost  
never reach ground state. The frozen length scale, however, is system-size dependent with a linear relationship. 
Furthermore, for the domain growth, unambiguous confirmation of the $t^{1/2}$ behavior has been provided. 
Results obtained by using Metropolis and Glauber algorithms are found to be consistent with each other. 
Here we state that at $T_{f}=0$ the trial moves that bring no change in the energy are customarily accepted 
with the probabilities $p=0$, $1/2$ or $1$, the latter two correspond, respectively, to the Glauber and 
Metropolis methods. For $p=0$, like the previous studies \cite{ole,ole2}, we observe frozen dynamics from very early 
time and these results are not presented.  
\par
The rest of the paper is organized as follows. In Section $2$ we discuss the methods. Results 
are presented in Section $3$. Finally we summarize our results in Section $4$.
\section{Methods}
Nonconserved coarsening dynamics in the nearest neighbor Ising model, introduced above, for $T_f=0$, 
is studied via MC simulations \cite{lan} in periodic square ($d=2$) and cubic ($d=3$) boxes. 
We have used square lattice in $d=2$ and simple cubic lattice in $d=3$. The values of $T_c$ for this model \cite{lan} 
in $d=2$ and $3$ are respectively $\simeq 2.269 J/k_{B}$ and $\simeq 4.51 J/k_{B}$, $k_{B}$ being the Boltzmann constant.
\par
We have used the Glauber spin-flip moves \cite{gla}, a standard method to introduce nonconserved dynamics. 
For a trial move, the sign of a randomly chosen spin is changed. The move is accepted if such a change 
lowers the energy of the system and rejected if the move increases the energy. For no energy change, one can use 
different probabilities $p$ ($>0$) for accepting the moves \cite{ole,ole2}. In this work, we have used $p=0.5$ 
and $1$, that correspond to Glauber \cite{lan,gla} and Metropolis \cite{lan} acceptance probabilities, respectively. 
\par
Time in our simulations was measured in units of MC steps (MCS) \cite{lan}, one step consisting of $L^d$ 
trial moves, $L$ being the linear dimension of a system (in units of the lattice constant). For the sake 
of convenience, in the rest of the paper, we set $k_{B}$, $J$ and the lattice constant to unity.
\par
For the calculation of length \cite{das}, we have identified the sizes, $\ell_d$, of various domains by 
scanning a system along different Cartesian directions. Two successive changes in sign in any direction 
identify a domain and the corresponding distance provides the length, i.e., the value of $\ell_d$. 
The average value, $\ell(t)$, was obtained from the first moment of the time-dependent distributions, 
$P(\ell_{d},t)$, thus obtained, i.e., 
\begin{equation}
 \ell(t)=\int{d\ell_d \ell_d P(\ell_{d},t)}.
\end{equation}
The value of $\ell$ can be obtained from the scaling properties \cite{bray,puri} of $C(r,t)$ and $S(k,t)$ 
as well [see Eqs. (\ref{e5_sclcor}) and (\ref{e5_sclsf})]. The measures from different functions are 
expected to provide the same information, apart from different constants of proportionality. Note that 
the spin variable $S_i$ is similar to the order parameter field $\psi$. Thus, with respect to the calculations of 
various correlation functions no further discussion becomes necessary.
\par
All our results are presented after averaging over multiple independent initial configurations. This number, 
for growth and aging, falls in the range between five and $60$, depending upon the value of $L$.  
Other than the finite-size effects and freezing phenomena related analyses, all data for the correlation functions 
are for $L=512$, presented after averaging over 100 and 20 independent initial configurations, in $d=2$ and $d=3$, 
respectively. For the freezing phenomena, given that the studied systems are rather small, we have obtained the 
quantitative results after averaging over several hundred initial configurations.
\begin{figure}
\centering
\includegraphics*[width=0.40\textwidth]{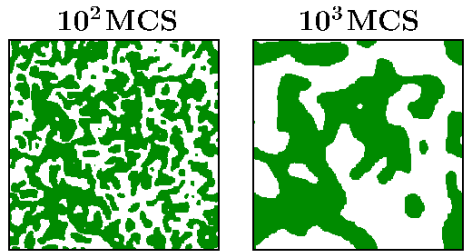}
\caption{\label{fig1} Evolution snapshots from the Monte Carlo simulations of the 2D nonconserved 
Ising model at $T_{f}=0$, after quenching from $T_i=\infty$. 
These pictures correspond to $p=1$. The marked regions represent ``up''
spins and the locations of the ``down'' spins are left unmarked. The linear dimension of 
the system is $L=512$.
}
\end{figure}
\section{Results}
First, we discuss results for $p=1$, in subsection $3.1$. Following this, in subsection $3.2$ we will present 
results for $p=1/2$.
\subsection{$p=1$}
\par
\begin{figure}
\centering
\includegraphics*[width=0.4\textwidth]{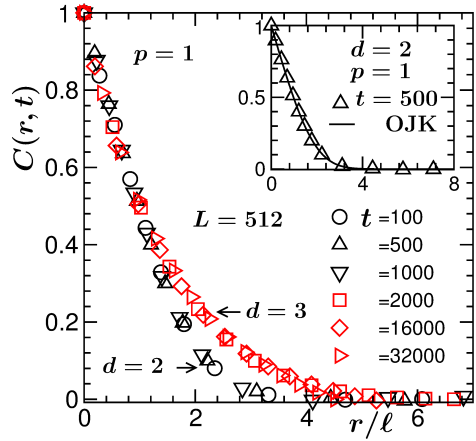}
\caption{\label{fig2}
Scaling plots of the two-point equal-time correlation function. We have shown $C(r,t)$ as 
a function of $r/\ell$, for both $d=2$ and $3$. In each dimension data from three 
different times are included. The used values of $\ell$ in this figure were obtained 
from $C(\ell,t)=0.5$. For all the other purposes we have used $\ell$ obtained 
from the first moment of the domain size distribution function $P(\ell_{d},t)$. 
Inset: Plot of $C(r,t)$ versus $r/\ell$ for $d=2$ and $t=500$. The continuous 
curve there is the Ohta-Jasnow-Kawasaki function. All results are for $p=1$.}
\end{figure}

For the sake of completeness, as well as for the convenience of later discussion, we start by presenting 
results for the pattern and growth. In Fig. \ref{fig1} we show snapshots taken during the evolution of the nonconserved 
Ising model at $T_{f}=0$. These snapshots are from MC simulations in $d=2$. Growth in the system is clearly 
visible. To check for the self-similarity, in Fig. \ref{fig2} we show scaling plots of the two-point 
equal-time correlation function. In this figure, we present data from both $d=2$ and $3$. Nice collapse of data, 
in both the dimensions, from different times, when plotted versus $r/\ell$, confirms self-similar 
character \cite{bray,puri} of the growth. Interestingly, the master curves from $d=2$ and $d=3$ do not match 
with each other \cite{das}. As shown in the inset, the $d=2$ data are in agreement with the OJK function 
\cite{ohta,das}. We note here, $C(r,t)$ at high temperatures, for both the dimensions, agree well 
with the OJK form \cite{das}. This states the fact that the pattern at $T_f=0$, in $d=3$, is special. 
This is in line with previous reports by other authors \cite{das,amar,shore,lipow,cue,cor2,ole,ole2,chak,chak_2} and will 
be useful in explaining new observation with respect to aging property.
\par
\begin{figure}
\centering
\includegraphics*[width=0.4\textwidth]{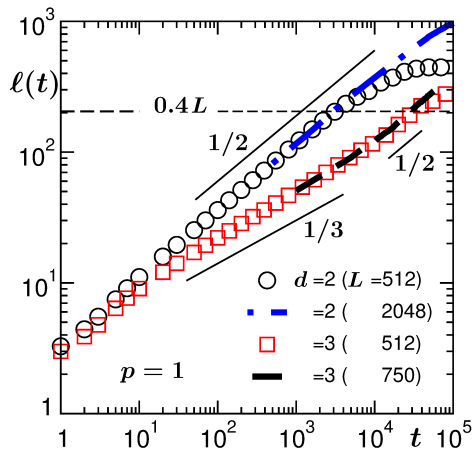} 
\caption{\label{fig3}Log-log plots of average domain size versus time, for $p=1$. We have shown data from both the 
dimensions. The continuous lines represent various power laws, exponents for which are 
mentioned in the figure. The dashed horizontal line, at $\ell=0.4L$, marks the appearance of 
finite-size effects for data sets corresponding to $L=512$.
}
\end{figure}

In Fig. \ref{fig3} we present data for domain growth, viz., we show $\ell(t)$ 
versus $t$, on a log-log scale. Results from both the dimensions are included. The 2D 
data exhibit a unique power-law behavior, with $\alpha \simeq 1/2$, over an extended 
period of time. The departure from the above scaling in the long time limit, clearly seen for $L=512$ (shown with symbols),  
is due to finite-size effects. While this behavior in $d=2$ is same as the results 
from nonzero temperature, the case of $d=3$ is very different from our observation 
for coarsening at $T_f$ higher than the roughening transition \cite{van} (here note that in 
$d=2$ there does not exist a nonzero roughening transition). The 3D data in Fig. \ref{fig3}, 
after a brief initial period (with higher exponent corresponding to annihilation of local defects), 
displays \cite{cor,das,amar,shore,lipow,cue} growth with $\alpha \simeq 1/3$ over more than two 
decades in time. At very late time the data exhibit a crossover \cite{das,cor2,chak} to $\alpha \simeq 1/2$. 
Towards the end of the presented time window, we see signature of finite-size effects for $L=512$ 
(see the data set with symbols). 
\par
With respect to the appearance of finite-size effects, however, there is 
similarity between the 2D and 3D cases. See the deviation of data with symbols, corresponding to $L=512$ for both the 
dimensions, at $\ell\simeq205\,(= 0.4 L)$, marked by the dashed horizontal line. 
For larger system size (see the dashed line for $d=3$ with $L=750$ and the dashed-dotted line for $d=2$ with $L=2048$), 
of course, the growth with 
$\alpha=1/2$ is more prominent due to lesser effects of finite system size over the presented time range. Note that such 
a clear confirmation of $t^{1/2}$ behavior for extended period of time in $d=3$ could not have been previously possible 
because of consideration of much smaller system sizes. The appearance of finite-size effects at $\ell\simeq0.4L$ at 
$T_f=0$ is very similar to that for non-zero values of $T_f$. We mention here, in the 3D case the very late time 
finite-size behavior for $T_f=0$ is rather complex \cite{ole,ole2}. The systems almost 
never reach the ground state (this problem is perhaps less severe in $d=2$) even at the end of extremely long simulation 
runs \cite{ole,ole2,spr,psen}. One may anticipate this fact to be somewhat more severe 
for $p<1$, which we will investigate later. Now we move to the aging property, the primary objective 
of the work.
\begin{figure}
\centering
\includegraphics*[width=0.4\textwidth]{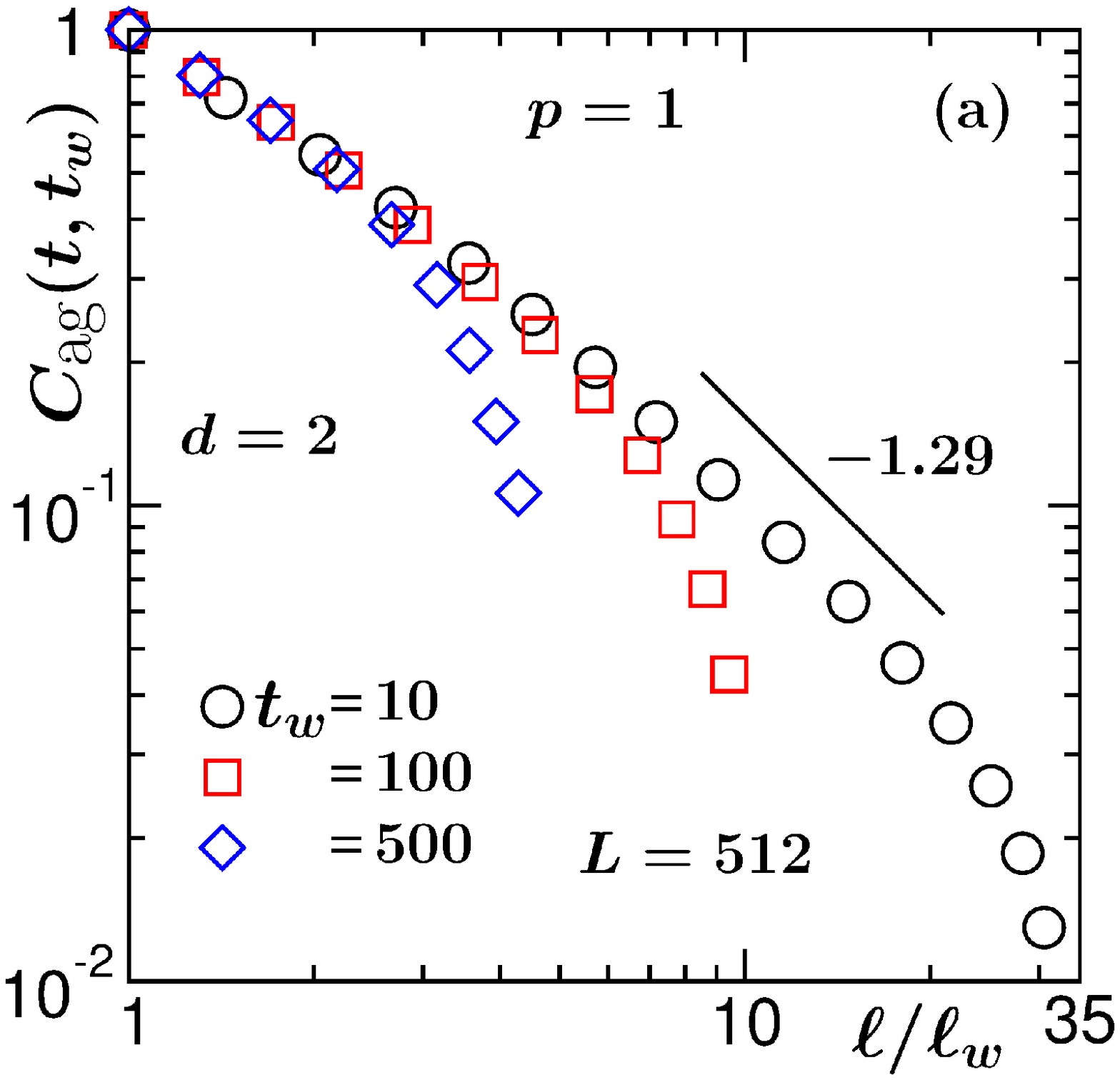}
\vskip 0.3cm
\includegraphics*[width=0.4\textwidth]{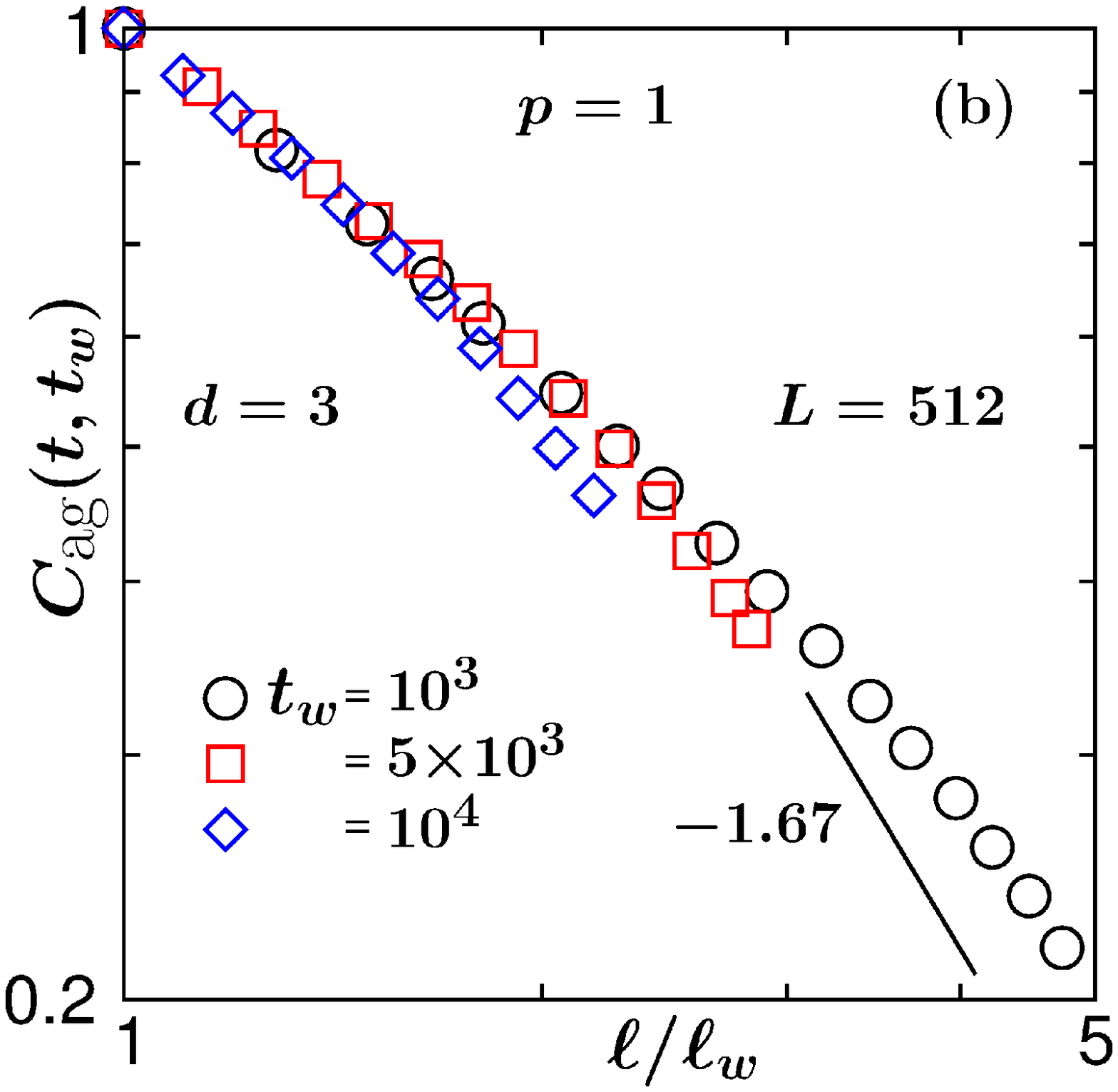}
\caption{\label{fig4}
(a) Log-log plots of the autocorrelation function, $C_{\textrm{ag}}(t,t_{w})$, 
vs $\ell/\ell_w$, for $d=2$. Results from a few different values of $t_w$ 
are shown. The solid line corresponds to a power-law decay with exponent $\lambda=1.29$. 
(b) Same as (a), but here the results are from $d=3$. The solid line here has the power-law exponent $\lambda=1.67$. 
All the results correspond to $p=1$.
}
\end{figure}
\par
In Figs. \ref{fig4}(a) and \ref{fig4}(b) we demonstrate the scaling property of the autocorrelation 
function \cite{das}. In these figures, we have plotted $C_{\textrm{ag}}(t,t_{w})$ versus   
$\ell/\ell_w$, for different values of $t_w$. Results in Fig. \ref{fig4}(a) are from 
$d=2$, whereas the $d=3$ data are presented in Fig. \ref{fig4}(b). Very nice 
collapse of data can be appreciated for both the dimensions. Deviations from the collapse, for large values of the 
abscissa variable, are related to finite-size effects \cite{mid1,mid2}. This occurs at a smaller value of $x$ for 
a larger choice of $t_w$, as expected. This is because, for a higher value of $t_w$ smaller fraction of the system 
size is available for further growth.
For $d=3$, the presented data did not suffer as much from the 
finite-size effects. In this dimension running simulations for large systems over very long period, to observe 
such effects, is computationally very demanding. Note that the appearance of finite-size effects in the autocorrelation 
function again complies with our above quoted limit $\ell\simeq 0.4 L$ that we observe for the domain growth. 
The solid lines in these figures are power-laws with LM values of 
$\lambda$. Clearly, there exist discrepancies between the LM exponents and the simulation results. 
Furthermore, we observe continuous bending \cite{mid1} in the scaling functions obtained from the simulations, on the 
log-log scale. This, of course, is possible when there exist corrections to the power-law scaling \cite{mid1}. 
In such a situation, calculation of the instantaneous exponent \cite{liu,mid1,mid2,maz,huse},
\begin{equation}\label{e5_lambdai}
\lambda_{i}=-\frac{d \ln C_{\textrm{ag}}}{d \ln x},
\end{equation}
can provide useful information. 
\begin{figure}
\centering
\includegraphics*[width=0.4\textwidth]{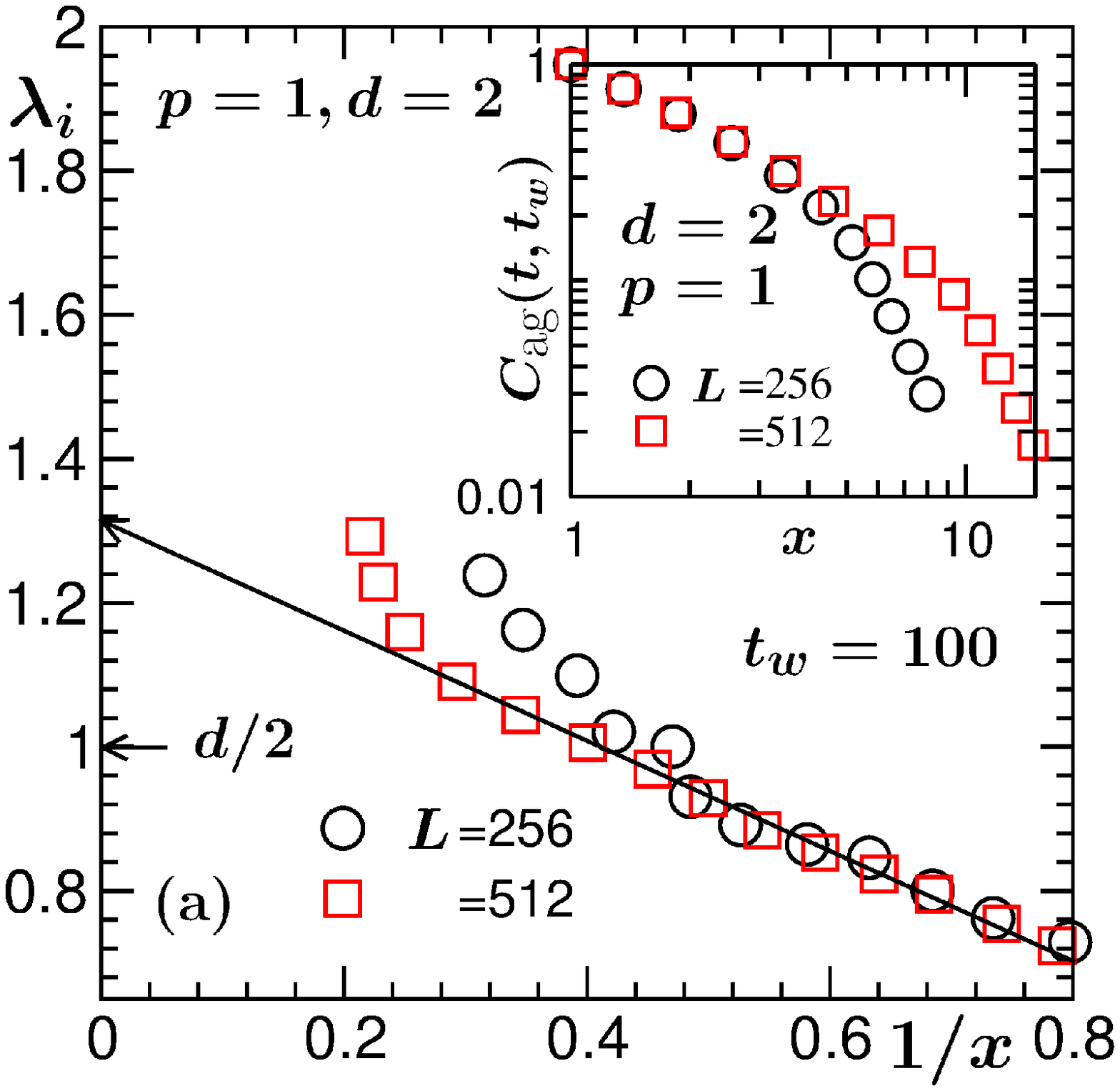}
\vskip 0.3cm
\includegraphics*[width=0.4\textwidth]{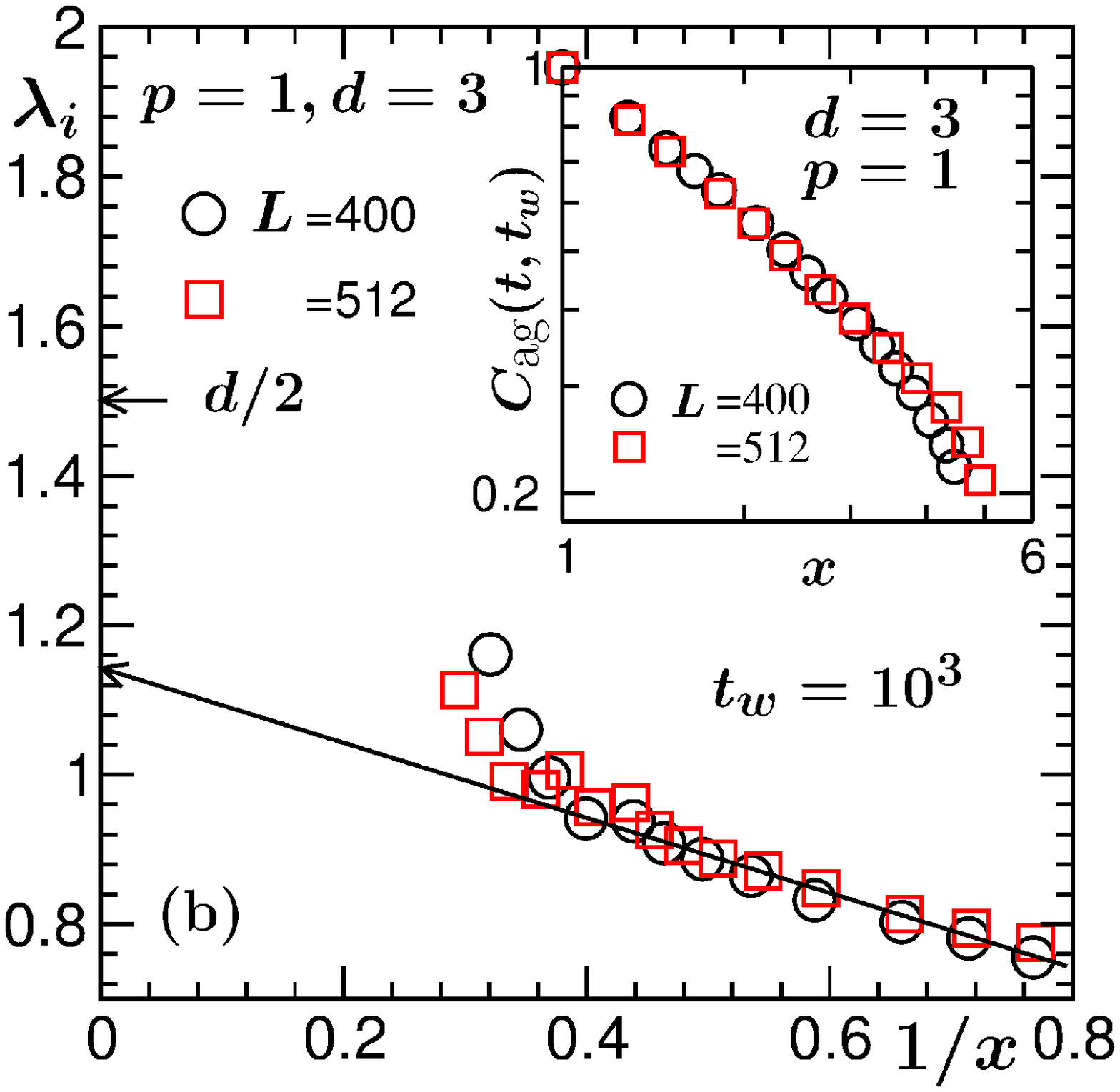}
\caption{\label{fig5}
(a) Instantaneous exponent, $\lambda_i$, for the $d=2$ Ising model, 
is plotted as a function of $\ell_w/\ell$, for different system sizes. The 
value of $t_w$ is mentioned on the figure. The solid line is a guide to the eye. 
(b) Same as (a), but the results here are from $d=3$. In both the figures the 
FH lower bounds have been marked by $d/2$. The insets in (a) and (b) show corresponding results for 
$C_{\textrm{ag}}(t,t_w)$ versus $\ell/\ell_w$, on log-log scale. All the results were obtained by 
fixing $p$ to $1$.
}
\end{figure}
\par
In Fig. \ref{fig5}(a) we show $\lambda_i$, as a function of $1/x$ (recall, $x=\ell/\ell_w$), for the $d=2$ case. 
Results for two different system sizes, for a fixed value 
of $t_w$, are included. Corresponding $C_{\textrm{ag}}(t,t_w)$ versus $x$ plots, on log-log scale, are shown in the inset. 
Data for the smaller value of $L$ deviate from a linear behavior, as $x$ increases, 
when the value of $x$ is approximately $2$. This is due to the finite-size effects and can be appreciated from the 
continued linear trend, up to much larger value of $x$, exhibited by the data from the 
larger value of $L$. Results for different values of $t_w$, for fixed $L$, provide similar information. 
This is, as stated above, because of lesser effective system size available for larger $t_w$.
\par
A linear extrapolation to $x= \infty$, using data unaffected by finite-size effects, 
provides a value of $\lambda$ close to $1.3$. Invoking the linear behavior,
\begin{equation}\label{e5_lambdai_linear}
 \lambda_i = \lambda -B/x,
\end{equation}
in the  definition 
in Eq. (\ref{e5_lambdai}), one obtains an exponential correction factor \cite{mid1,mid2}, i.e., 
\begin{equation}\label{e5_aging_full}
C_{\textrm{ag}}(t,t_{w})=Ae^{-B/x}x^{-\lambda},
\end{equation}
$A$ and $B$ being constants.
This implies, a power law can be realized only in the $t >> t_w$ limit. Like in the critical phenomena here also it is 
natural to expect that the corrections should be described by power-laws. However, separately estimating exponents for 
corrections of different orders is a difficult task. Given the trend of the data sets, the exponential factor appears 
to describe the corrections reasonably well. We will make further comment on the accuracy of this full form later. 
\par
Similar linear behavior is observable in Fig. \ref{fig5}(b) where we have presented data for  
$d=3$ (again, for corresponding $C_{\textrm{ag}}(t,t_w)$ versus $\ell/\ell_w$ plots see the inset). 
In this case, the data exhibit convergence 
to a value \cite{chak_2} between $1.1$ and $1.2$. While for $d=2$ the convergence is consistent with that for higher 
temperature \cite{mid1}, there exists serious departure in the case 
of $d=3$ from the LM prediction. Note that the LM prediction in $d=3$ matches well with the 
conclusions from the simulation studies at higher temperatures, above the roughening transition \cite{mid1}. 
Furthermore, $\lambda\simeq1.15$ is far below the lower bound of FH. 
\par
Even though reasonably accurate estimate is possible from such extrapolations, one can do better by performing 
finite-size scaling analysis \cite{mid1,mid2}, given that the data for $\lambda_i$, at large $x$, may suffer 
from statistical error and 
finite-size effects, preventing unambiguous choice of regions for performing a linear fit. 
A finite-size data collapse exercise will be further useful for 
bringing confidence in the form of Eq. (\ref{e5_aging_full}), which essentially is an 
empirical form.
\begin{figure}
\centering
\includegraphics*[width=0.4\textwidth]{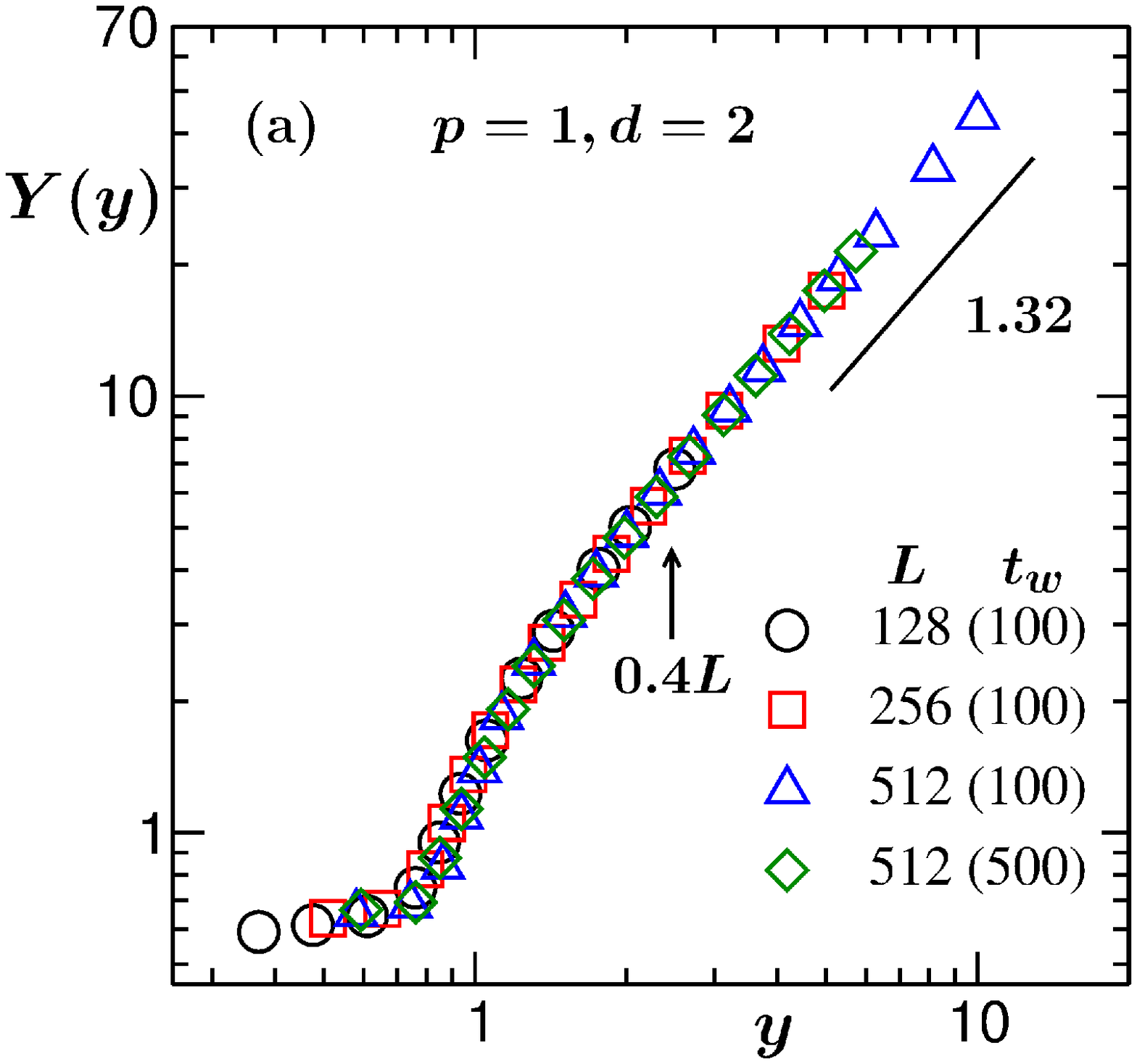}
\vskip 0.3cm
\includegraphics*[width=0.4\textwidth]{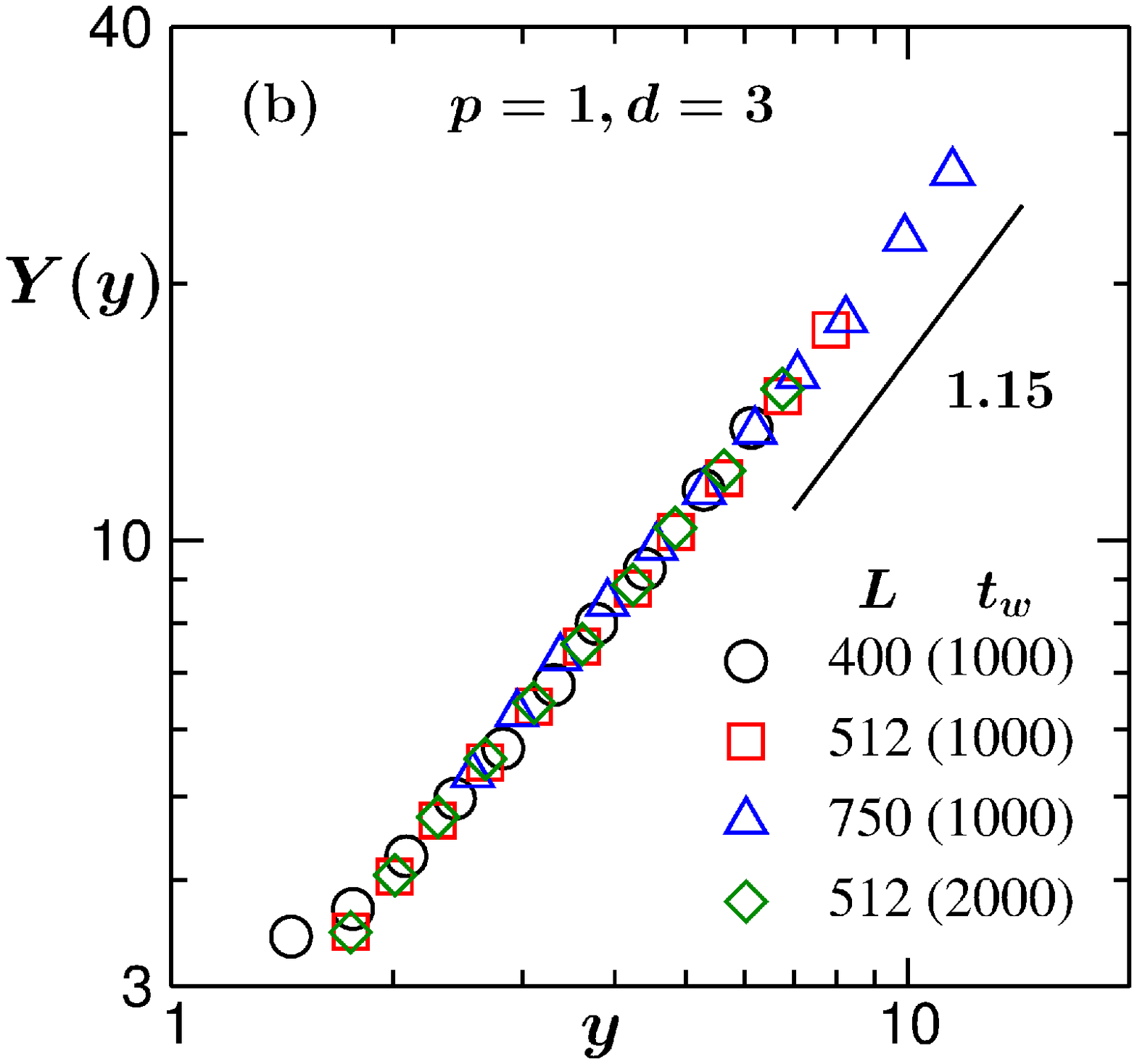}
\caption{\label{fig6}
(a) Finite-size scaling analysis of the autocorrelation function in $d=2$. The vertical arrow marks the departure 
of the scaling function from the $y^{\lambda}$ behavior. 
(b) Same as (a), but here it is for $d=3$. The solid lines represent power-laws with the values of exponent 
mentioned next to them. All results are for $p=1$.
}
\end{figure}
\par
In a finite-size scaling method one looks 
for collapse of data from various system sizes \cite{lan,fish,fis_3}. Such a 
method for the analysis of the data for autocorrelation function was recently constructed \cite{mid1,mid2}. 
See Refs. \cite{maz2} for more recent work with a different system.
Like in the critical phenomena \cite{lan,fish,fis_3}, here also one introduces a scaling function $Y$. To make $Y$ 
independent of system size, we search for a dimensionless scaling variable $y$. Since $x\,(=\ell/\ell_w)$ is already 
dimensionless, we choose $y=x'/x$, where $x'=L/\ell_w$. This provides
\begin{equation}
 y=\frac{L}{\ell}.
\end{equation}
\par
For the sake of convenience, we intend to write $C_{\textrm{ag}}(t,t_w)$ as a function of $y$. Then,
\begin{equation}\label{e5_fss}
C_{\textrm{ag}}(t,t_{w})=Ae^{-By/y_w}\left(\frac{y_w}{y}\right)^{-\lambda},
\end{equation}
where $y_w=L/\ell_w$, i.e., the value of $y$ at $t=t_w$. Next, we write the finite-size scaling 
function, a bridge between thermodynamic and finite-size limit behavior, as
\begin{equation}
 Y=C_{\textrm{ag}}(t,t_{w})e^{By/y_w}y_w^{\lambda}.
\end{equation}
In the above equation we have absorbed $y^{\lambda}$ inside $Y$. 
\par
We stress again that the choice of $x'$ above is driven 
by the dimension of $x$ and the fraction of the total system size 
available to explore, given that the measurement starts at $t_w$. 
Coming back to the above comment ``fraction of the total system size available to explore'', we mention 
that this quoted fact allows a finite-size scaling analysis only via the variation of $t_w$, without 
exploring different system sizes. This is because, we state again, with the variation of $t_w$, the above mentioned 
fraction varies, providing different effective system sizes. This fact we will demonstrate by achieving 
collapse of data from different values of $L$ and $t_w$.
\par
The limiting behavior of $Y$ can be described as follows. In the thermodynamic limit, i.e., 
for $\ell << L$ $(y\rightarrow \infty)$, 
we can write
\begin{equation}
 Y=Ay^{\lambda},
\end{equation}
so that Eq. (\ref{e5_fss}) is recovered. In the other limit, in finite systems, particularly due to frozen 
dynamics, we do not expect 
$C_{\textrm{ag}}(t,t_w)$ to vanish. This may lead to a rather flat appearance of $Y$ for small $y$. 
Such characteristic features, 
along with a collapse of data from various different $L$ and $t_w$ values, can 
be realized if $\lambda$ is chosen appropriately, alongside the constant $B$. 
In our data collapse exercise we will treat these two quantities as adjustable 
parameters. 
\par
At nonzero temperatures, there exists coupling between equilibration 
of domain magnetization and that of the whole system \cite{puri}, till large value of $x$. Given that 
the former is related to the 
critical fluctuation \cite{fis_2}, for very low value of $T_f$, the relaxation 
related to the domain magnetization occurs very fast, to a value almost unity. Nevertheless, a minor jump in 
the autocorrelation function very close to $x=1$, providing a higher effective exponent for very small $x$, exists. 
Thus, we avoid the data point corresponding to $x=1$ in all cases, for the 
finite-size scaling analysis.  
Furthermore, scaling of $C_{\textrm{ag}}$, with respect 
to $\ell/\ell_w$, is expected to be observed from rather small values of $t_w$, very small $T_f$. 
Nevertheless, deviations at early time is observed, particularly in $d=3$. 
This may be due to slow crossover to $t^{1/2}$ growth behavior 
extending up to very late time. Thus, for this scaling analysis, we have chosen rather 
large values of $t_w$ in this dimension. 
\par
Even though in an earlier study \cite{mid1} (for high temperatures) we have obtained 
good data collapse by using finite-size $\ell$ in the scaling variable $y$, ideally one should use the  
thermodynamic limit values. There can be two possible ways: (i) to adopt $\ell\sim t^{1/2}$ behavior, 
(ii) to use length from a much larger system size that does not exhibit finite-size effects over the time-scale of 
analysis. We follow the latter method here (as well as for $p=1/2$) -- for $d=2$, $\ell$ will be taken 
from $L=2048$ and for $d=3$, we will 
use $\ell$ from $L=750$. 
\par
Results from the finite-size scaling analysis for $d=2$ are presented in Fig. \ref{fig6} (a), whereas 
corresponding results for $d=3$ are presented in Fig. \ref{fig6} (b). In the 
case of $d=2$, very good collapse of data, along with consistency with the limiting behavior discussed above, 
is obtained for $\lambda=1.32$ 
and $B=0.80$. This value of $\lambda$, within 
statistical error, is in agreement with a previous study \cite{mid1} for $T_f=0.6T_c$ 
and consistent with the prediction of LM. In the large $y$ limit the data are consistent with the power-law amplitude 
$A\simeq2$. This can be appreciated by considering $C_{\textrm{ag}}(t,t_w)=1$ at $x=1$ and $B=0.8$. 
The departure from the 
power-law behavior is marked by a vertical arrow in Fig. \ref{fig6} (a). This corresponds to $\ell=0.4L$ where 
finite-size effects occur. A robust power law behavior for $Y$ till the finite-size effects appear, irrespective of 
the system size, provide confidence in the exponential correction factor. In this connection, also note that for 
$L=512$ and $t_w=100$, the power-law behavior extends over $t-t_w$ ranging between $0$ and approximately $3600$. 
One may think of improving accuracy in the 
estimation of $\lambda$ by allowing for an adjustable exponent in Eq. (\ref{e5_lambdai_linear}), by replacing $x$ by 
$x^\gamma$. This exponent will, of course, appear in the argument of the exponential factor. 
We caution here that the scaling analysis will be 
less reliable if $\gamma$ is included, due to large number of adjustable parameters. 
The information with respect to the departure point of the scaling 
function from the power-law is quantitatively similar in other cases as well. So, we will not discuss it again.
\par
For $d=3$, on the other hand, the 
number for the exponent $\lambda$ turns out to be $1.15$, which is very different from that at $0.6T_c$ \cite{mid1}. 
Again note that the high temperature result in this dimension is in good agreement with 
the LM value. Furthermore, the value of $\lambda$ at $T_{f}=0$ is far below the 
lower bound of FH and this conclusion is consistent with that from the analysis of 
the instantaneous exponent. The question then comes, is it a 
true violation of the bound? This can perhaps be understood from the derivation of YRD. Before moving to that discussion 
we briefly point out  
the intermediate range features of $Y$ in Fig. \ref{fig6}. Before exhibiting a nearly flat behavior the 2D data 
fall rather sharply, compared to the 3D case. This may be related to the more prominent freezing phenomena in 
the latter dimension. This fact, as promised, we will discuss later.
\begin{figure}
\centering
\includegraphics*[width=0.4\textwidth]{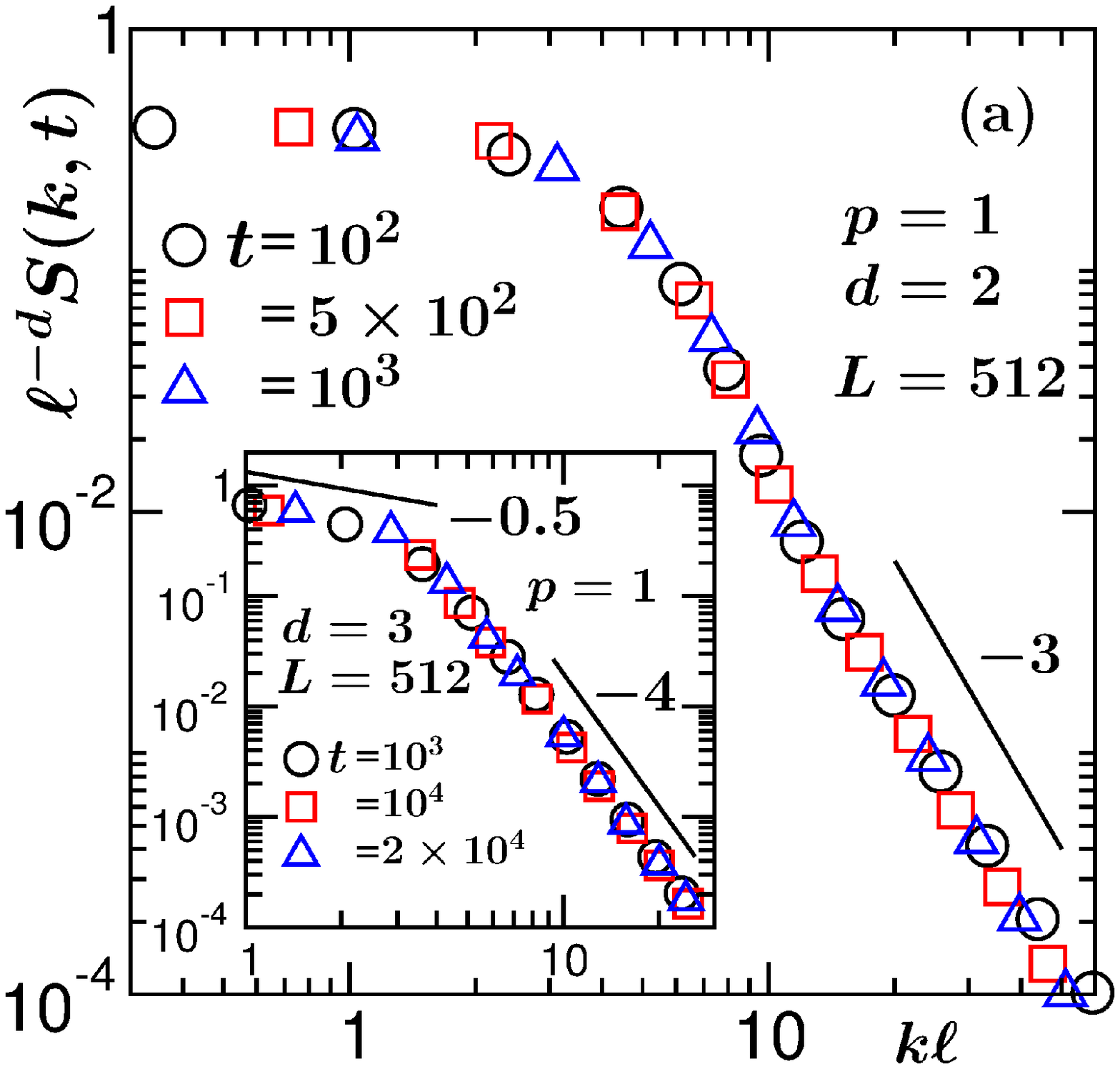}
\vskip 0.4cm
\includegraphics*[width=0.43\textwidth]{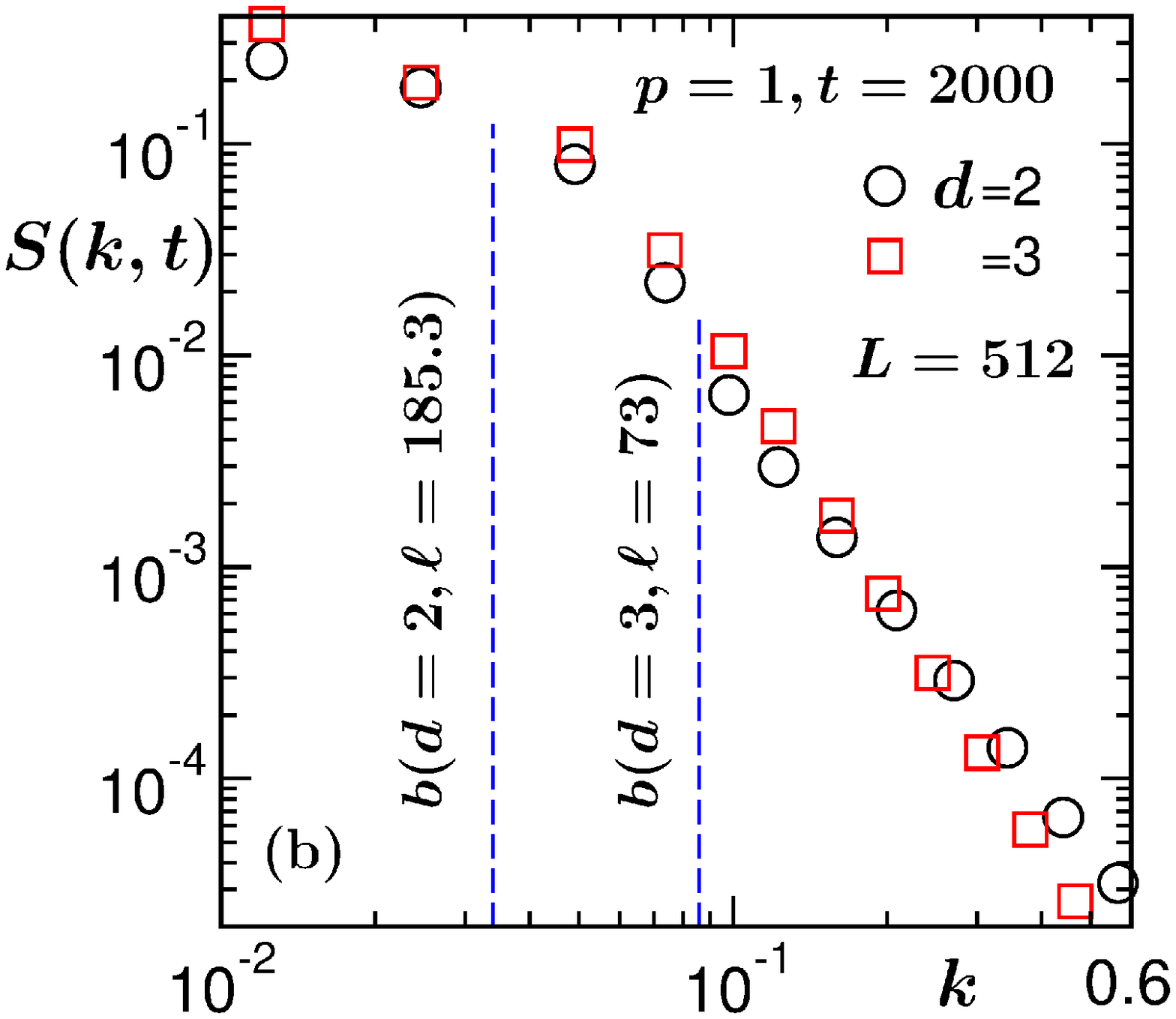}
\caption{\label{fig7}
(a) Scaling plots of the structure factor in $d=2$ (main frame) and $d=3$ (inset). 
The solid lines are power-laws, exponents for which are mentioned in the figure. 
(b) Plots of the structure factor, vs $k$, on a double-log scale, 
for $d=2$ and $d=3$. The dashed vertical lines are explained in the text. These results 
are for $p=1$.
}
\end{figure}
\par
In Fig. \ref{fig7}(a) we show the plots of $S(k,t)$ from 
$d=2$ (main frame) and $3$ (inset). Our focus here is to obtain the 
scaling behavior of Eq. (\ref{e5_sclsf}). Nice collapse of data, in 
both the dimensions, signify that the chosen values of $t_w$ for the 
finite-size scaling analyses are well inside the scaling regime of structure and growth. The 
power-laws with exponent $-3$ and $-4$ represent the Porod law \cite{porod,oono}
in $d=2$ and $3$, respectively. 
\par
Using the equal time structure factors at $t_w$ and $t$, YRD arrived at
\begin{equation}\label{e5_swartz}
C_{\textrm{ag}} \leq \ell^{d/2}\int_{0}^{b}{dk k^{d-1} [{S(k,t_{w})} {\tilde{S}(k\ell)}]^{1/2}},
\end{equation}
in which they substituted the small $k$ behavior of $S(k,t_{w})$ 
[cf. Eq. (\ref{e5_small_wave})], to obtain the lower bound. In Eq. (\ref{e5_swartz}),
$b$, the upper limit of the integration, equals $2\pi/\ell$. 
In Fig. \ref{fig7}(b) we present $S(k,t)$ vs $k$ plots on a 
log-log scale from $d=2$ and $3$. For both the dimensions we have chosen $t=2000$ MCS. 
The vertical dashed lines there correspond to the upper limit $b$ 
for different dimensions, corresponding to a time ($10^4$ MCS) 
reasonably larger than $2000$ MCS. It appears, the upper limit of integration for $d=3$ 
covers a significant range of $k$ over which $S(k,t)$ decreases, 
providing an ``effective'' negative value of $\beta$.
Given that the growth of $\ell$ is much slower in $d=3$ than in $d=2$ over a long 
intermediate period, one needs to go to very long time to access $\beta=0$ behavior. 
Even then, notice that a negative value of $\beta$ can already be appreciated 
from small $k$ regime in the inset of Fig. \ref{fig7}(a). Thus, even if 
a crossover occurs to the value predicted by LM, 
simulations with much larger systems with orders of magnitude 
longer period of time will be needed to observe that. For that purpose, a change 
in pattern may also be necessary at such late time.
\begin{figure}
\centering
\includegraphics[width=0.4\textwidth]{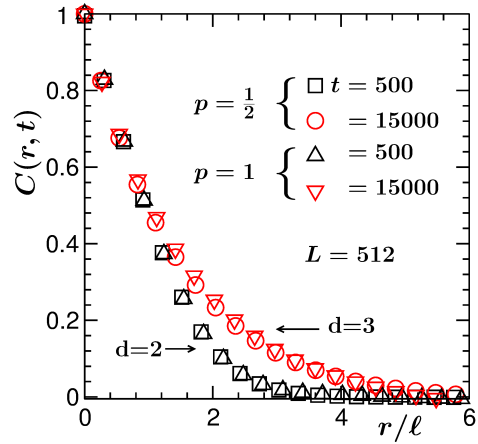} 
\caption{\label{fig8}
Two-point equal-time correlation functions, from MC simulations 
of the nonconserved Ising model in $d=2$ and $3$, obtained by using 
$p=1/2$ and $1$, are compared. The distance axis is scaled 
by the average domain length $\ell$. Like in Fig. \ref{fig2}, here also 
the value of $\ell$ was obtained as the distance at which $C(r,t)$ decays 
to half its maximum value.
}
\end{figure}
\begin{figure}
\centering
\includegraphics[width=0.4\textwidth]{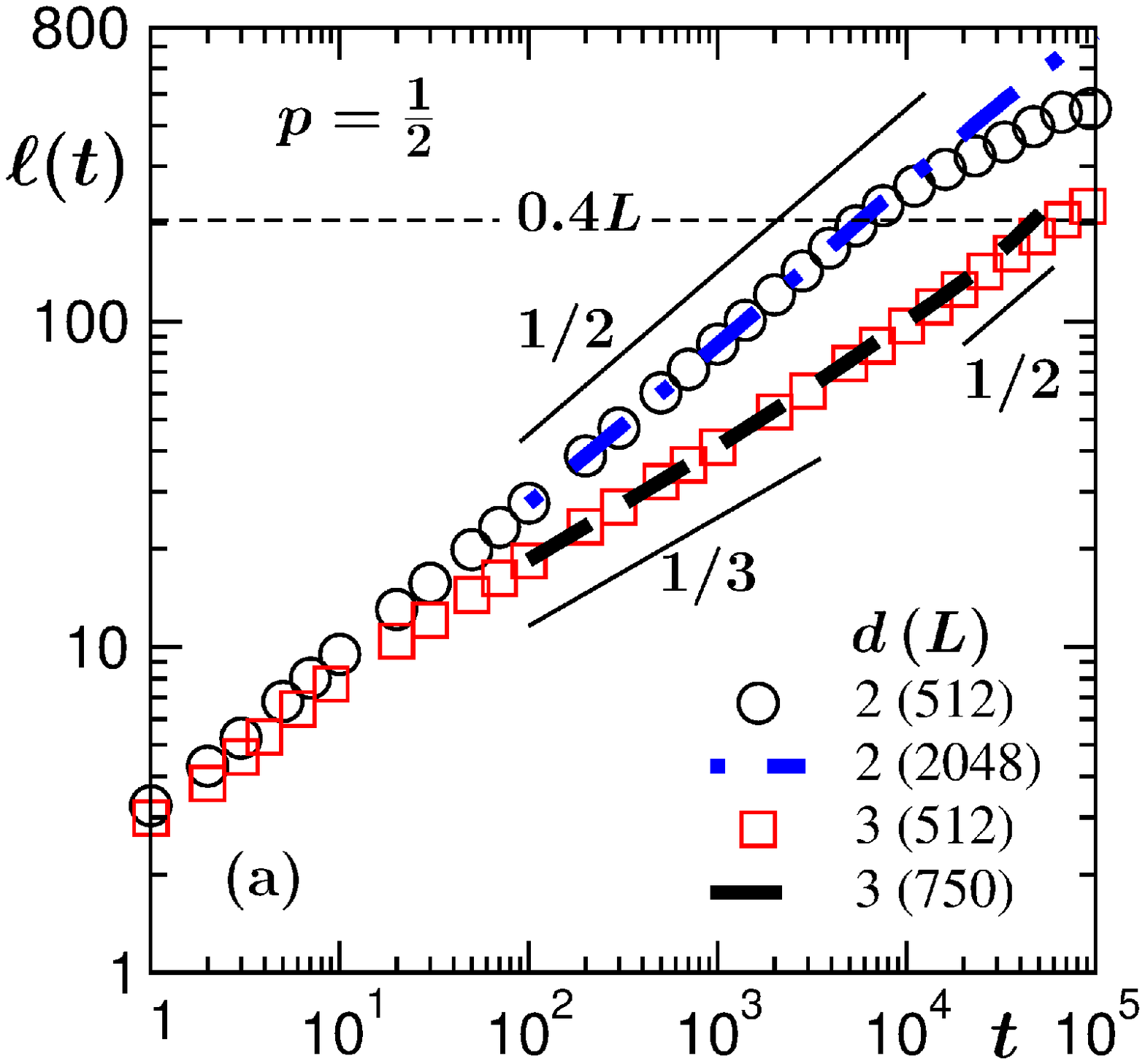}
\vskip 0.3cm
\includegraphics[width=0.4\textwidth]{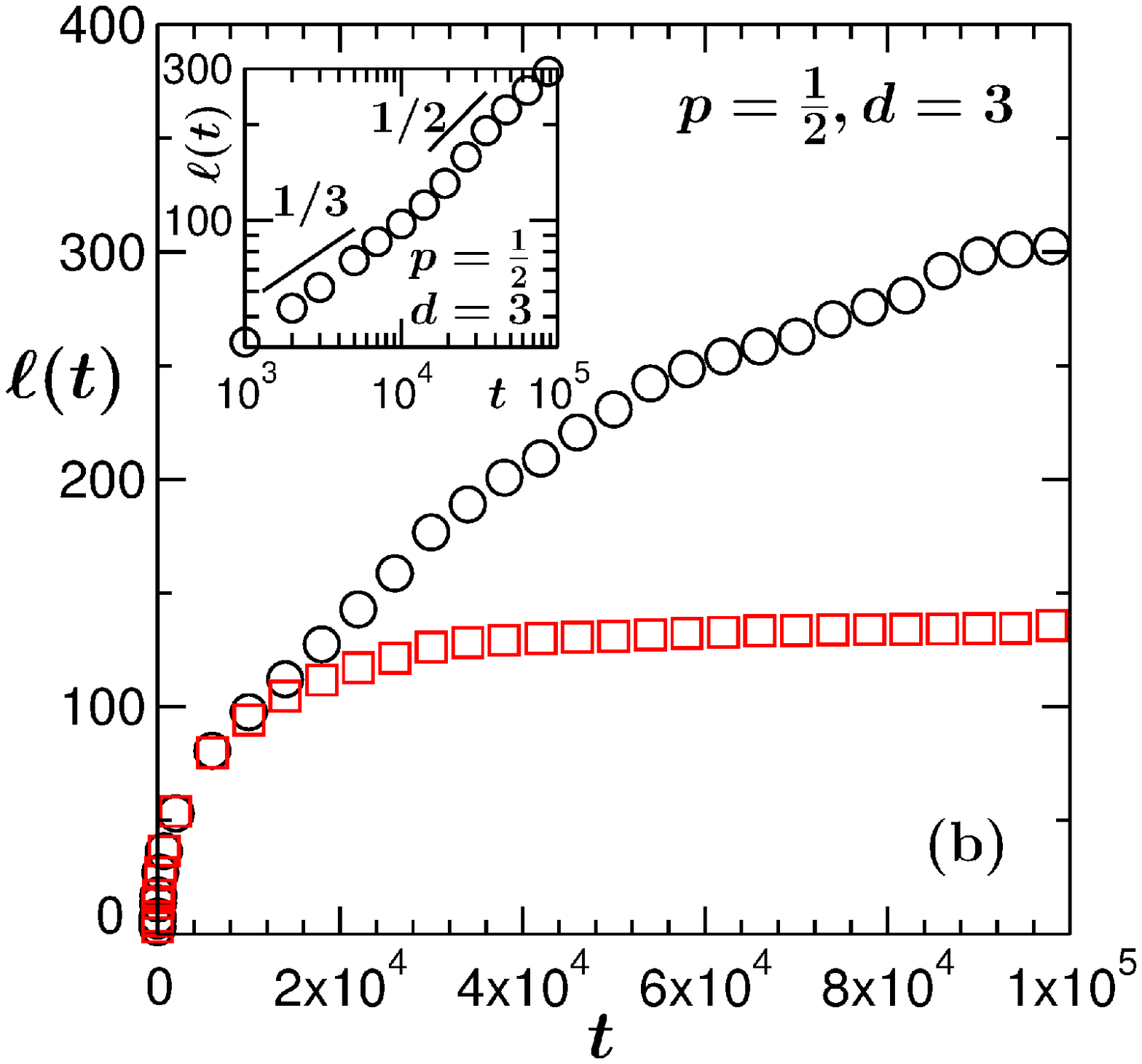}
\caption{\label{fig9}
(a) Log-log plot of average domain length versus time. We have shown data 
from $d=2$ and $3$. The continuous lines 
represent power-laws, exponents being mentioned next to them. The dashed horizontal line marks the 
appearance of finite-size effects, at $\ell=0.4L$, for $L=512$. 
(b) Plot of domain length versus time in $d=3$, for two different initial 
configurations. Inset shows a log-log plot of average domain size versus time 
for the data set that did not undergo freezing. These results are obtained 
with $p=1/2$.  
}
\end{figure}
\begin{figure}
\centering
\includegraphics[width=0.4\textwidth]{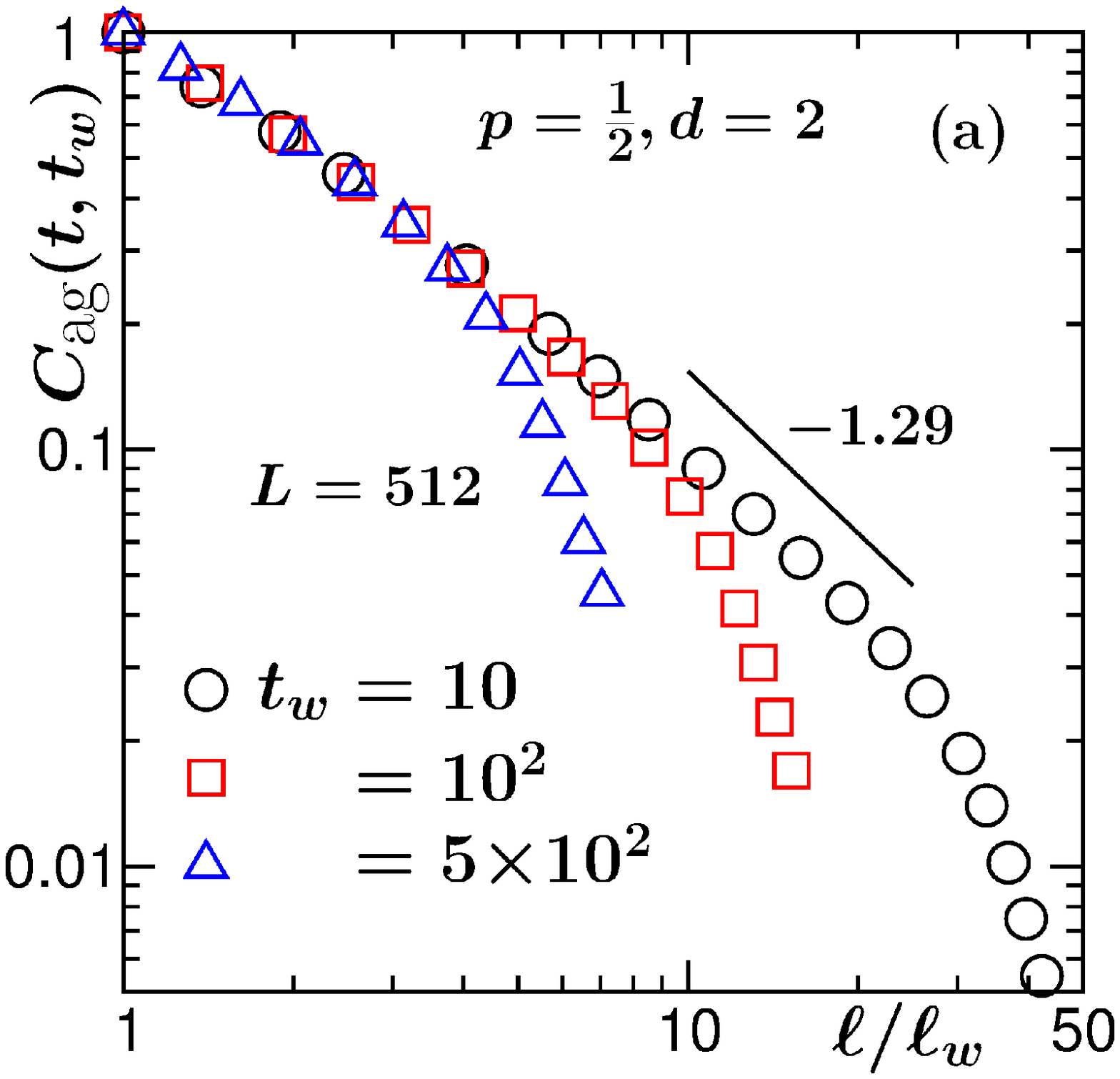} 
\vskip 0.3cm
\includegraphics[width=0.4\textwidth]{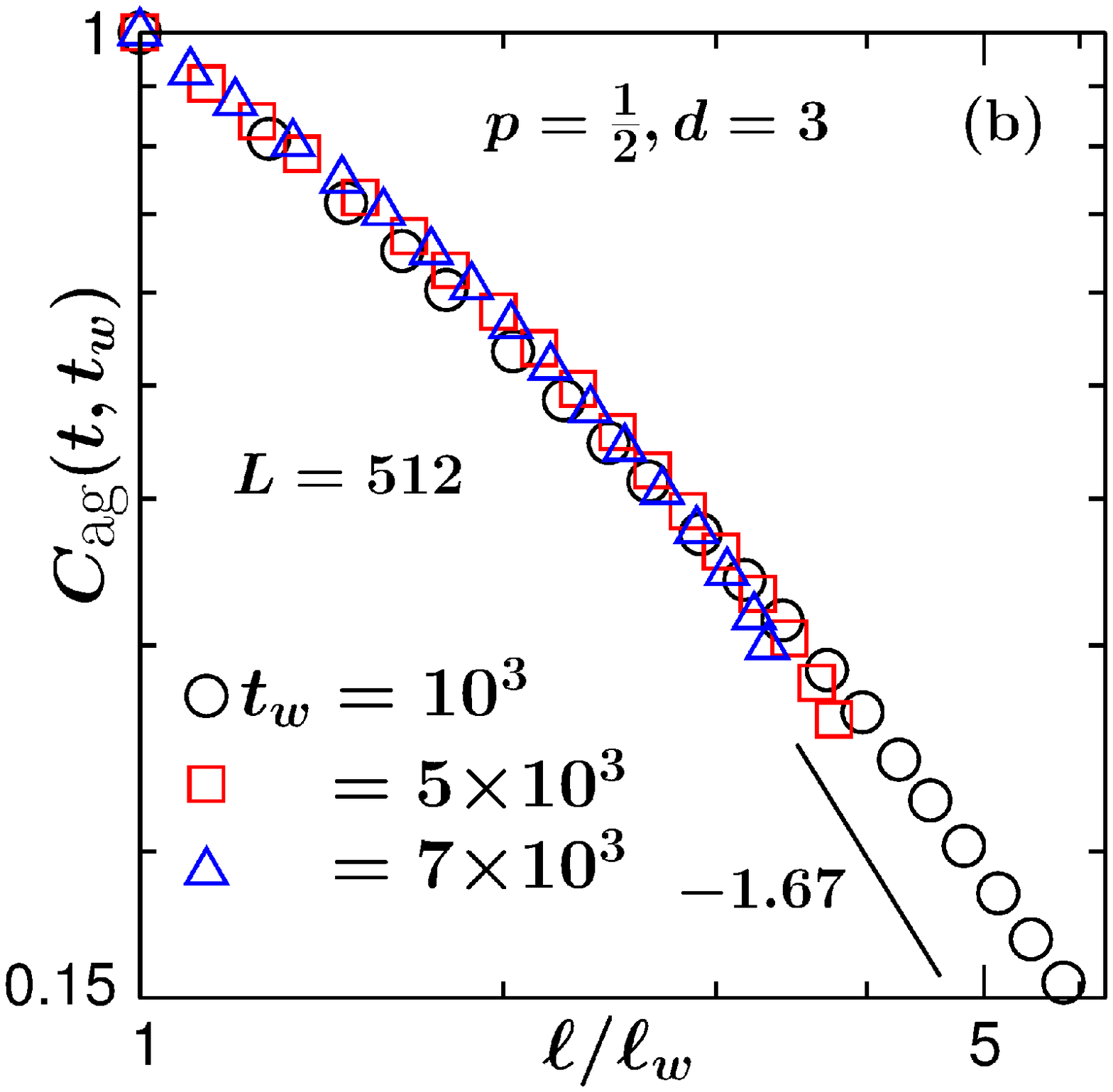}
\caption{\label{fig10} (a) Two-time autocorrelation functions, from $d=2$, for different 
values of $t_w$, are plotted versus $\ell/\ell_w$, on a log-log scale. 
(b) Same as (a) but for $d=3$. The solid lines represent power-laws, 
exponents for which are mentioned in the figures. All results correspond to 
$p=1/2$.
}
\end{figure}
\begin{figure}
\centering
\includegraphics[width=0.38\textwidth]{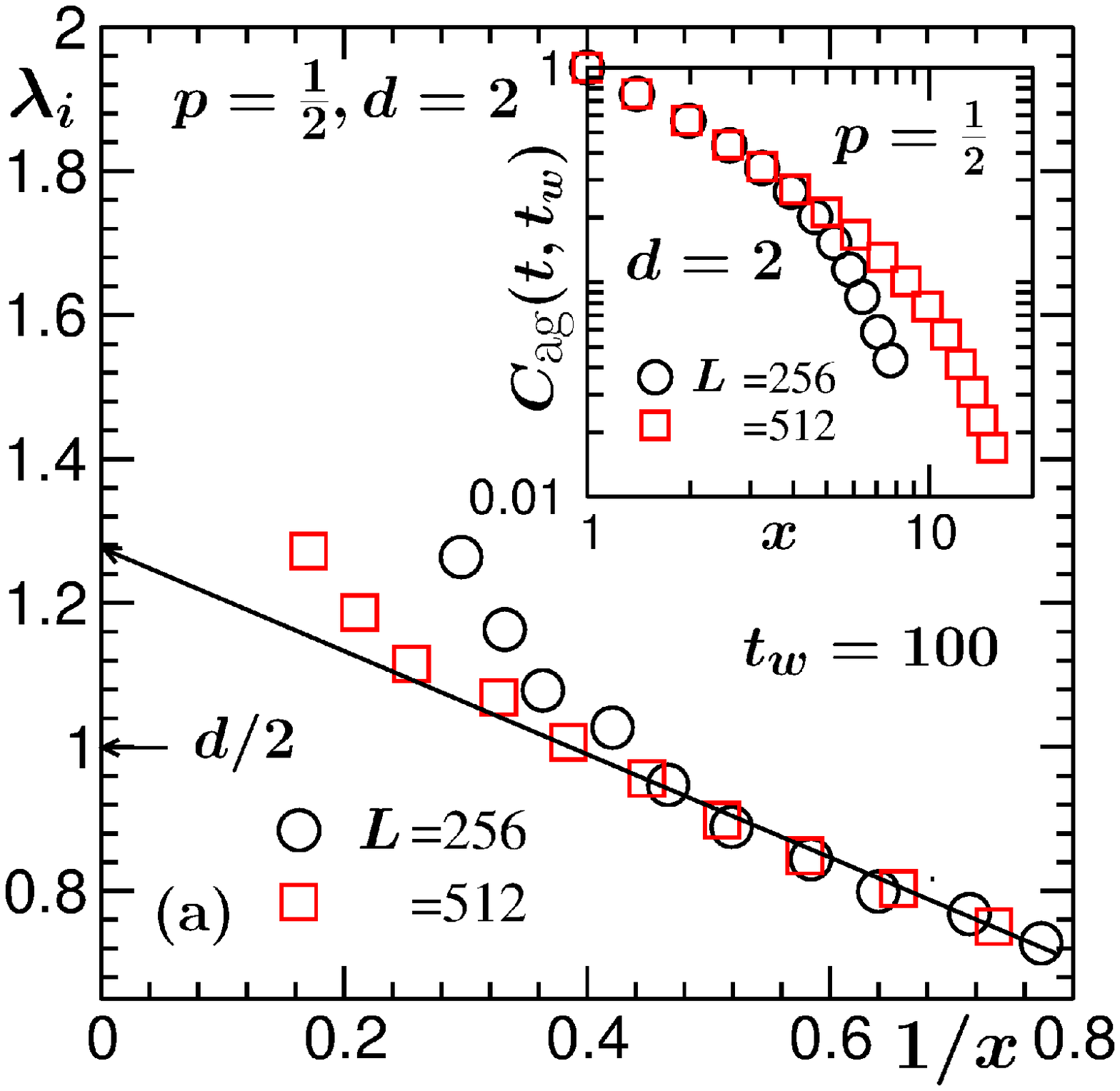} 
\vskip 0.3cm
\includegraphics[width=0.38\textwidth]{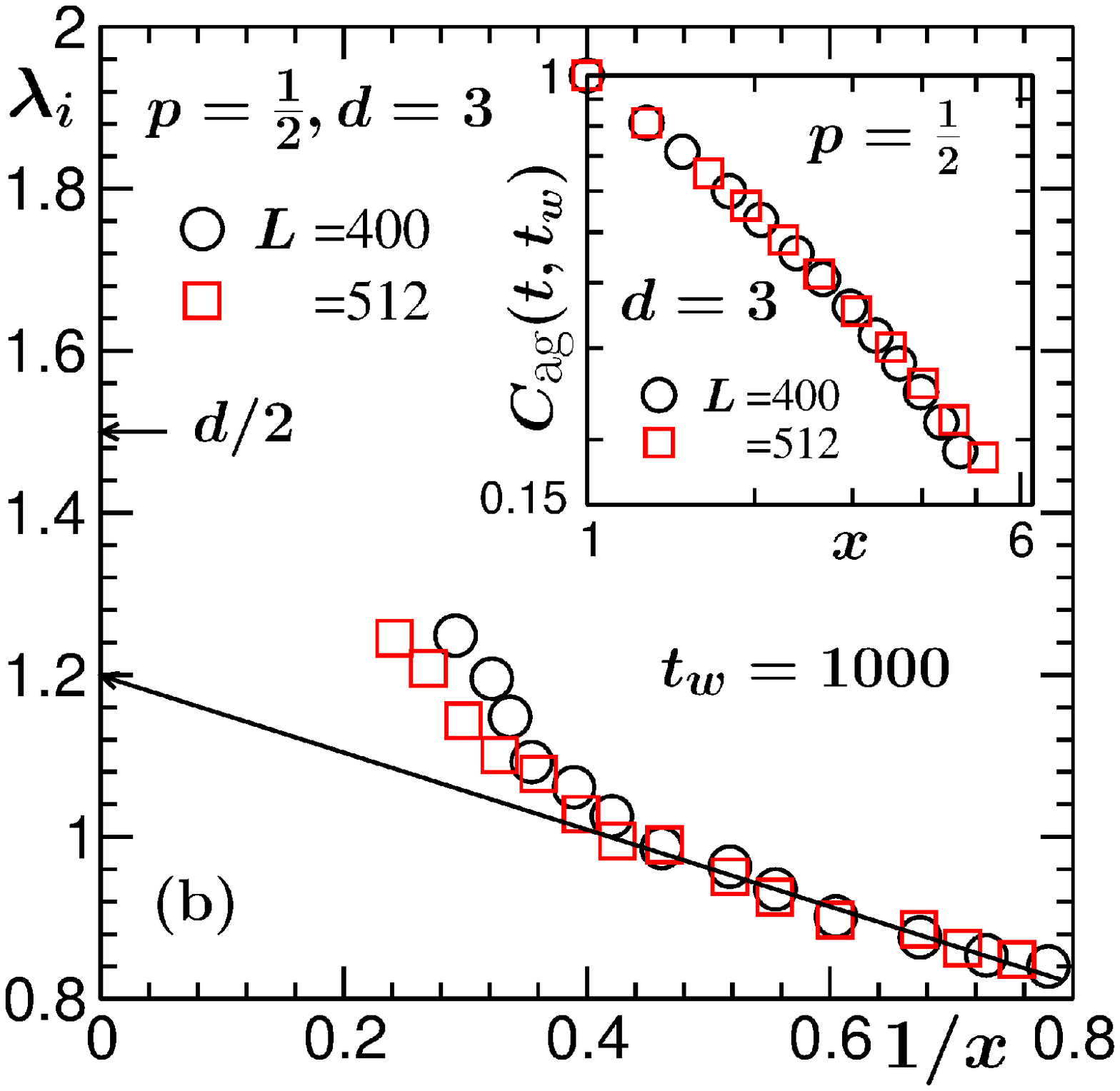}
\caption{\label{fig11} 
Plots of instantaneous exponent $\lambda_i$ versus $1/x$ ($x=\ell/\ell_w$), 
from (a) $d=2$ and (b) $3$. We have fixed the value of $t_w$, and in each dimension, presented 
results from two different system sizes. The solid straight 
lines are guides to the eye. The FH lower bounds have been marked by $d/2$ in both (a) and (b). 
Corresponding direct plots, i.e., data for $C_\textrm{ag}(t,tw)$ versus $x(=\ell/\ell_w)$, are shown 
in the insets. These results are from simulations with $p=1/2$.}
\end{figure}
\par
As stated earlier, an objective of the paper is to compare MC simulation results with different algorithms, 
viz., we intend to check if there are any differences in the outcomes when trial moves bringing no energy change 
are accepted with probabilities \cite{ole,ole2,lan} $p=1$ and $1/2$. Having presented the results for $p=1$, next 
we focus our attention to the case with $p=1/2$, with occasional direct comparison of the results with those for 
$p=1$, wherever seem important.
\subsection{$p=1/2$}
\par
In Fig. \ref{fig8} we present comparative results for the two-point equal-time correlation function. This 
figure contains data from both the dimensions, for $p=1/2$ as well as for $p=1$. 
It appears that the results for $p=1/2$ nicely overlap with 
those for $p=1$ when the distance axis is appropriately scaled by the  corresponding average domain sizes. 
These results confirm that both the algorithms provide similar structure.
\par
Fig. \ref{fig9} (a) shows log-log plots of $\ell$ versus $t$, for both the dimensions. 
Here we have included data only from the $p=1/2$ case. Like in the $p=1$ case, the 2D data appear consistent with the 
theoretical expectation all the way till the finite-size effects appear \cite{das} at $\ell \simeq 0.4 L$. 
For $L=512$, see the departure of the data set from that of $L=2048$. Plot for 
the $d=3$ case also exhibits a trend similar to the $p=1$ case -- there exists a consistency of the data set with 
an exponent $\alpha\simeq 1/3$ for nearly three decades in time, after which a crossover appears. This is very clearly 
visible in the $L=750$ case. By examining the $L=512$ data one may conclude that the finite-size effects appear a little 
earlier than when $\ell$ reaches $0.4L$. This could well be due to statistical reasons. 
Here note that runs for many initial configurations get trapped in metastable states very early, without allowing us to
appropriately probe the post-crossover region. Data presented in Fig. \ref{fig9} (a) are averaged by including such 
runs as well.  
In Fig. \ref{fig9} (b) we show $\ell$ versus $t$ data for two typical 
runs, on linear scale. It is clear that some runs can encounter freezing around the time (or even before) 
when a crossover is expected. This necessitates either extremely good statistics or very large system size. 
In the inset of this figure, we show the plot for the data set that did not show a signature of 
freezing, on a log-log scale. Clearly, a $t^{1/2}$ behavior is very prominent towards the end. This overall picture 
is true for both values of $p$.  We will provide further discussion on freezing phenomena 
towards the end of this subsection. Next we present 
results for aging.
\par
Figs. \ref{fig10} (a) and \ref{fig10} (b) are similar to that of Figs. \ref{fig4} (a) and \ref{fig4} (b), except for the 
fact that here we present autocorrelation function for $p=1/2$. Nice collapse of data from different $t_{w}$ values are 
visible when plotted versus $\ell/\ell_{w}$. The runs for $d=3$ [Fig. \ref{fig10} (b)] are not long enough to 
see the finite-size effects as clearly as in the $d=2$ case [Fig. \ref{fig10} (a)]. Once again, even though there 
exist bending in the log-log plot, data in both the 
dimensions appear to get linear at late time, implying power-law decay with correction for finite $x$. 
While the 2D results appear reasonably 
consistent with the LM exponent, strong discrepancy can be seen in the case of $d=3$ (see comparison of the 
simulation data with the solid lines). These facts are consistent 
with our observation for $p=1$. 
\begin{figure}
\centering
\includegraphics[width=0.4\textwidth]{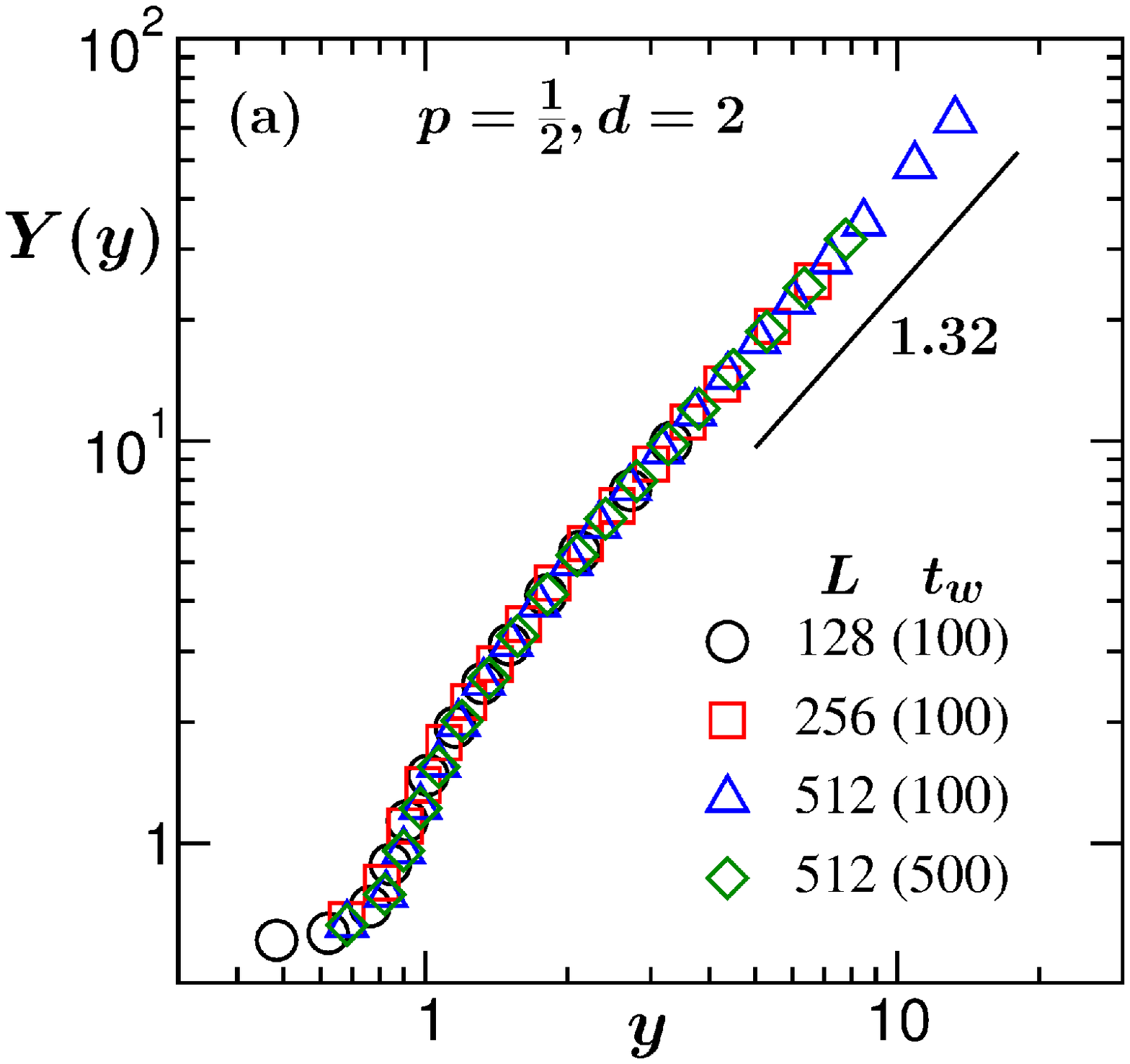} 
\vskip 0.3cm
\includegraphics[width=0.4\textwidth]{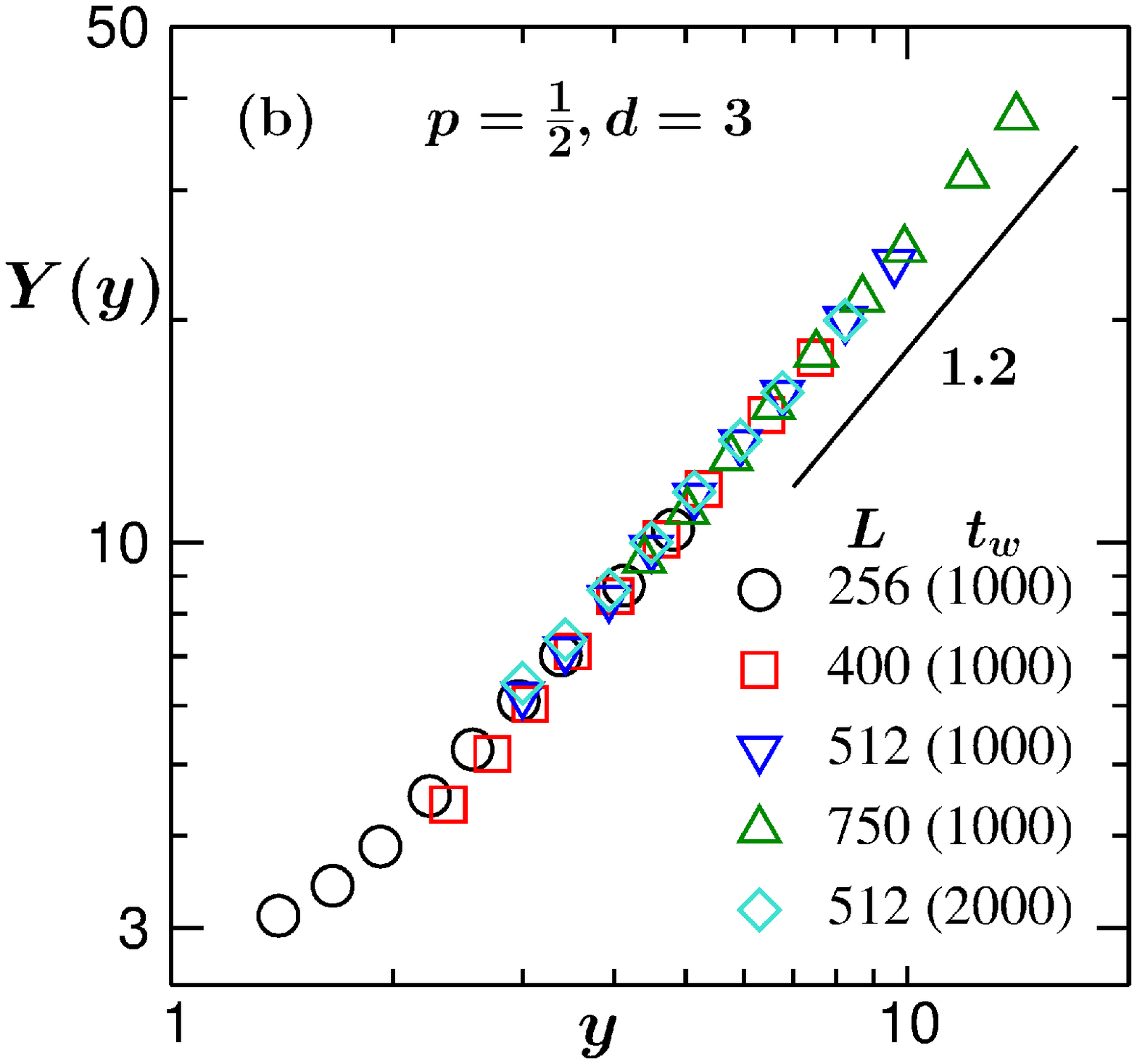}
\caption{\label{fig12}(a) Finite-size scaling analysis of the autocorrelation function in $d=2$. 
(b) Same as (a) but for $d=3$. Both the plots are for $p=1/2$. The solid lines represent power-laws with 
exponents $\lambda=1.32$ and $\lambda=1.2$ in $d=2$ and $3$, respectively.
}
\end{figure}
\par
To quantify the power-law exponents in the asymptotic limit, we calculate the instantaneous exponent 
\cite{mid1,mid2,huse} $\lambda_i$. 
The corresponding results for $d=2$ and $3$, vs $1/x$, are plotted in Figs. \ref{fig11} (a) and \ref{fig11} (b), 
respectively (see the corresponding $C_{\textrm{ag}}(t,t_w)$ versus $\ell/\ell_w$ plots in the insets), for different 
system sizes. The finite-size behavior appear consistent with the previous subsection. Furthermore, the data 
sets show convergence to values that are also consistent with the $p=1$ case (for comparison see Fig. \ref{fig5}). 
This is further confirmed by the finite-size scaling analysis plots which are presented in 
Fig. \ref{fig12} [see (a) for $d=2$ and (b) for $d=3$]. This exercise, as stressed 
earlier, discards any possibility of large errors 
that may appear due to finite size, statistical fluctuation or freezing effects. From the analysis we quote 
$\lambda\simeq 1.32$ in $d=2$ and $\lambda\simeq 1.2$ in $d=3$. Given that the two-point 
equal-time correlation functions from the two algorithms match nicely (see Fig. \ref{fig8}) with each other, viz, for 
$p=1/2$ and $p=1$, 
we do not take the route of explaining the violation of the FH bound in $d=3$ via the structure factor again. 
Rather, below we focus on the issue of freezing.
\par
The reason for discussing freezing is the following. Typically, in many real physical situations, the 
final (frozen) length scale, $\ell_f$, 
does not depend upon the system size \cite{das3,das4,mani,tung}. 
In many systems the value of $\ell_f$ is set by a distance related to the repulsive barrier in the 
interacting potential \cite{das3,das4,mani}. Such a picture related to barrier, however, is not expected in 
the current situation. Nevertheless, to rule out that none of the presented results are affected by this freezing phenomena, 
we need to know if there exists any system-size dependence of $\ell_f$.
\par
\begin{figure}
\centering
\includegraphics[width=0.3\textwidth]{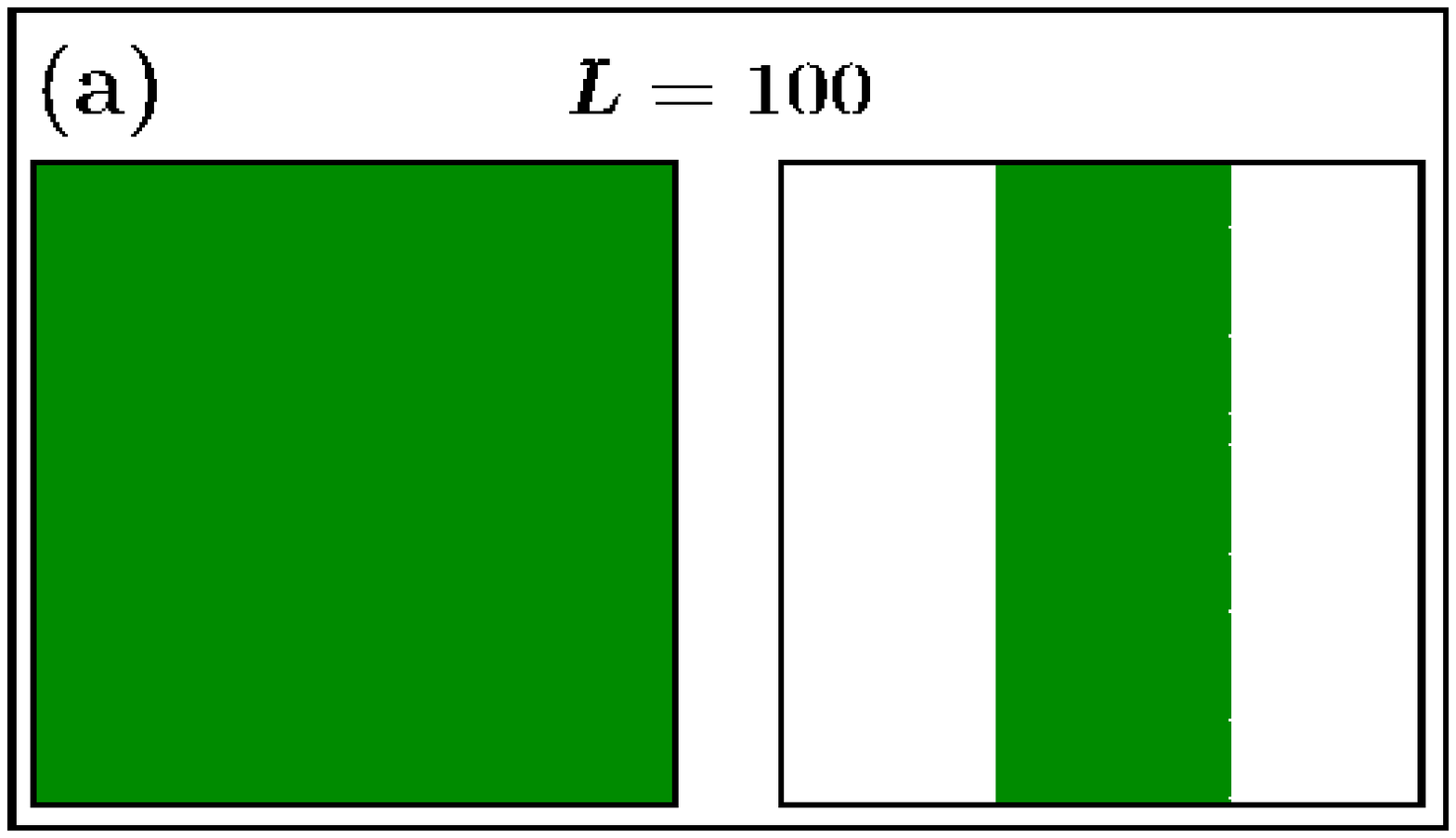} 
\vskip 0.3cm
\includegraphics[width=0.3\textwidth]{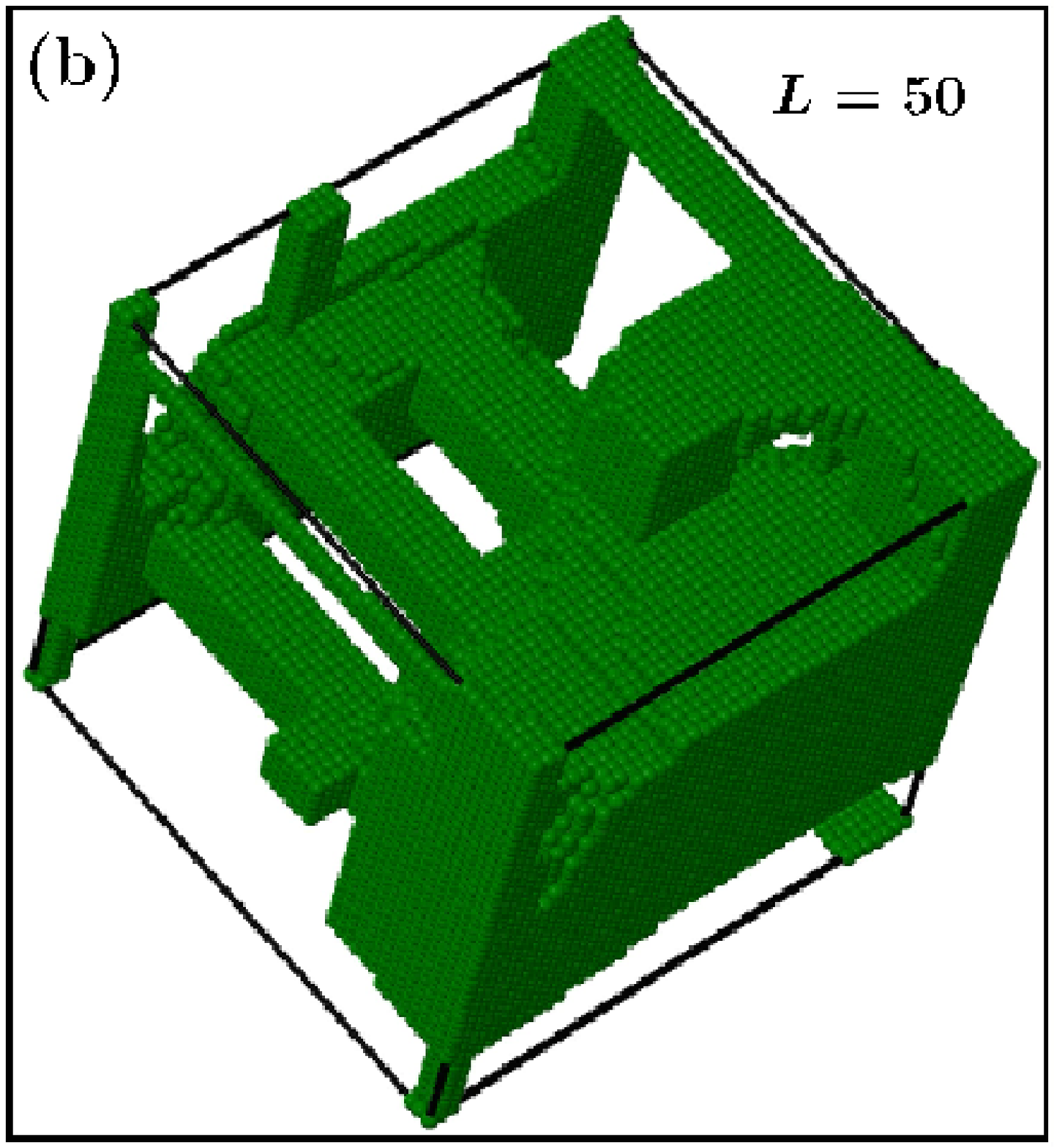}
\caption{\label{fig13} Typical final snapshots from representative configurations in (a) $d=2$ and (b) $b=3$. 
These snapshots were obtained by using $p=1/2$.
}
\end{figure}

In Fig. \ref{fig13} (a) we show final configurations from two different initial random compositions in $d=2$. 
The one on the left 
corresponds to the ground state and the other represents a frozen state. In Fig. \ref{fig13} (b) we show a frozen 
configuration from $d=3$. For both $d=2$ and $3$, the presented configurations are obtained by using $p=1/2$. 
As opposed to the right frame in $d=2$, where it is easily identifiable that no further growth can occur, $d=3$ 
structure is more 
complex. Details on this can be found in Refs. \cite{ole,ole2,chak_2,ole_3}.
\par
\begin{figure}
\centering
\includegraphics[width=0.45\textwidth]{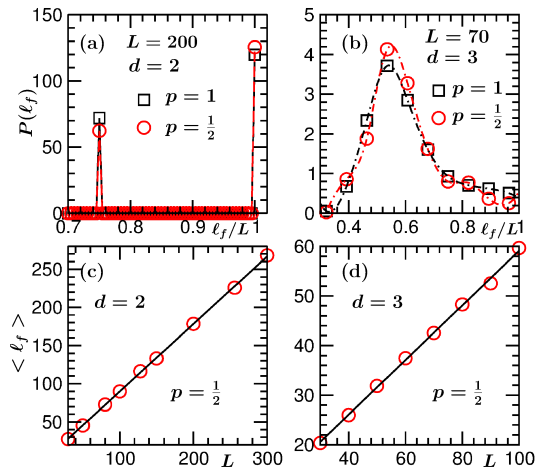}
\caption{\label{fig14} 
Plots of the distribution of final length, $\ell_f$, versus the scaled length $\ell_f/L$, for (a) $d=2$, $L=200$; 
(b) $d=3$ and $L=70$.  
Average value of the final length is plotted versus system size, for $p=1/2$ in (c) $d=2$ and (d) $d=3$. 
The solid lines represent linear fits to the simulation data sets.
}
\end{figure}

We observe that for a particular system size, for different initial configurations, $\ell_f$ varies significantly in $d=3$, 
almost never reaching $L$ \cite{ole,ole2}. Whereas in $d=2$, the frozen states are related to the stripe structure seen 
in Fig. \ref{fig13} (a). 
Distributions of $\ell_f$, $P(\ell_f))$, for both $p=1/2$ and $1$, 
obtained from such variation, are shown in 
Fig. \ref{fig14} (a), for $d=2$, $L=200$; and in Fig. \ref{fig14} (b), for $d=3$, $L=70$. 
The average values, $<\ell_f>$, that can be extracted from these 
distributions turn out to be approximately same for  
$p=1$ and $p=1/2$. The observation is similar for other values of $L$. In the case of $d=3$ we observe a single peak, with 
reasonably large width, 
whereas there exist two spikes in $d=2$. These spikes are related to the ground and stripe states (we have ignored the 
diagonal stripes). From the height of 
the two peaks, it can be appreciated that the ground state is reached approximately twice as often as the stripe 
states \cite{ole_3}. 
This fact also states that freezing is a more severe problem in $d=3$ than in $d=2$. 
\par
In Fig. \ref{fig14} (c) and \ref{fig14}(d) we plot $<\ell_f>$ as a function of $L$, for $p=1/2$, in $d=2$ and $3$, 
respectively. The dependence in both 
the dimensions turns out to be 
linear. Here it is worth mentioning that occasionally artificial freezing in computer simulations can be observed, 
for slow dynamics, due to periodicity 
in random numbers, when system sizes are very large. To us, this also does not appear to be true in the present case. 
Here the phenomena can be attributed to the structure. How the system size at this temperature is affecting the 
structure and dynamics, to provide a linear relation between frozen length and system size, is an intriguing question. 
The observation, nevertheless, provides confidence, by looking at the numbers in the plots [Figs. \ref{fig14}(c) and 
\ref{fig14} (d)], that our 
presented results on growth and aging did not 
suffer from this effect. Here we recall that finite-size effects start appearing when $\ell\simeq 0.4 L$, whereas 
$<\ell_f>$ in both the dimensions are much larger than this limit. 
Furthermore, linear dependence of $<\ell_f>$ on $L$, does not call for reanalysis of data via finite-size scaling 
method by replacing $L$ by $<\ell_f>$. 
Note that in Figs. \ref{fig14} (c) and \ref{fig14} (d) we presented results only up to $L=300$ and $100$, respectively, 
by considering 
the fact that achieving freezing for very large systems is computationally very difficult, particularly when our 
observation suggests a linear relationship.
\section{Conclusion}
We have studied pattern, growth and aging properties of the nearest neighbor ferromagnetic Ising model 
via Monte Carlo simulations \cite{lan}, using 
Glauber spin-flip \cite{lan,gla} moves. Flips which did not change energy 
were accepted with two probabilities, viz. $p=1$ and $1/2$, referred to as the 
Metropolis and Glauber algorithms, respectively, 
to check for their relative effects on the behavior of various quantities. 
Our focus was on zero temperature quench, for  both $d=2$ 
and $3$. Quantitative information on the decay 
of the two time autocorrelation function was 
obtained via finite-size scaling \cite{mid1,mid2}
and other methods of analysis. These were discussed with reference to the corresponding 
results for quenches to nonzero temperatures \cite{mid1}.
\par
The autocorrelations exhibit nice scaling with respect to $x$ ($=\ell/\ell_w$). The late 
time behavior is described by power-laws, $C_{\textrm{ag}}(t,t_{w}) \sim x^{-\lambda}$. At early time there exists 
correction that can be reasonably well described by an exponential factor. These features are very much similar to those for high temperature 
quenches \cite{mid1}.
\par
In $d=2$, the value of $\lambda$ is in agreement with the LM value.  
However, the $d=3$ result differs significantly from 
the high temperature result \cite{mid1}, the latter being consistent with the LM value. 
The estimated value in this dimension not only 
differs from the LM prediction, but also appears to be far below the lower-bound of FH. 
We argue, via analysis of the structure factor, in line with the 
derivation of YRD, that this is not a true violation if the small $k$ 
behavior of $S(k,t)$ is appropriately accounted for. Here note that the zero temperature structure in $d=3$ is 
incompatible with the well known Ohta-Jasnow-Kawasaki form. 
\par
As expected, all results 
are found to be nearly independent of aforementioned flipping probabilities. This is also true for 
the freezing property. For the latter we have demonstrated that the corresponding 
average length scale varies linearly with the system size. Close to the frozen length, the dynamics is very slow for $d=3$.  
If an analysis is performed, to arrive at the domain growth law, via scaling of the corresponding relaxation 
time with the system size, a different, misleading conclusion can be arrived at. Recall that, from simulations of very 
large systems we confirmed that in this dimension also the growth at zero temperature follows the Lifshitz-Allen-Cahn law. 
We will address this issue of very late time dynamics in a future 
communication. Before closing, we mention that despite no dependence of growth exponent on the acceptance probability $p$, 
the growth in the $p=1/2$ case is slower than when $p$ is set to unity, due to higher amplitude in the latter case. 
This fact is true, as expected, in both the space dimensions. Our observation suggests that finite-size effects also 
appear at a slightly smaller characteristic length scale for $p=1/2$. 


\end{document}